\documentclass[12pt]{article}
\usepackage{natbib}
\usepackage{apptools}
\usepackage{booktabs}
\AtAppendix{\counterwithin{lemma1}{section}}

\usepackage[]{appendix}

\newcommand{\samplestd}{\sigma}

\newcommand{\samplevar}{\samplestd^2}
\newcommand{\hatsamplevar}{\widehat{\samplestd}^2}

\usepackage{amsfonts,amsmath,amssymb,dsfont}       % blackboard math symbols
\usepackage{mathtools}
\usepackage{float}
\usepackage{subcaption}
\usepackage{xcolor}
\usepackage{hyperref}
\usepackage{xr}
\usepackage{booktabs}
\usepackage{array}
\usepackage{longtable}
\usepackage{multirow}

\def\E{\mathbb{E}}
\def\V{\mathrm{Var}}

\def\Pr{\mathrm{Pr}}
\def\AV{\mathrm{AVar}}

\DeclareMathOperator*{\argmin}{arg\,min}

\newcommand\indep{\protect\mathpalette{\protect\independenT}{\perp}}
\def\independenT#1#2{\mathrel{\rlap{$#1#2$}\mkern2mu{#1#2}}}

\newtheorem{thmlem}{Lemma}

\newtheorem{thmthm}{Theorem}

\newtheorem{suppthmlem}{Lemma}[subsection]

\newtheorem{suppthmcond}{Condition}[subsection]

\makeatletter

\newcommand*{\addFileDependency}[1]{% argument=file name and extension
\typeout{(#1)}% latexmk will find this if $recorder=0
\@addtofilelist{#1}
\IfFileExists{#1}{}{\typeout{No file #1.}}
}\makeatother

\newcommand*{\myexternaldocument}[1]{%
\externaldocument{#1}%
\addFileDependency{#1.tex}%
\addFileDependency{#1.aux}%
}
\newcommand{\blind}{1}
\myexternaldocument{supplementary}

\begin{document}

\def\spacingset#1{\renewcommand{\baselinestretch}%
{#1}\small\normalsize} \spacingset{1}

%%%%%%%%%%%%%%%%%%%%%%%%%%%%%%%%%%%%%%%%%%%%%%%%%%%%%%%%%%%%%%%%%%%%%%%%%%%%%%

\if1\blind
{
  \title{\large{\textbf{Robust integration of external control data in randomized trials}}}

\author{\small{Rickard Karlsson$^{1,\dagger*}$, 
Guanbo Wang$^{2,3\dagger*}$, Piersilvio De Bartolomeis$^{4}$, Jesse H. Krijthe$^{1}$, 
Issa J. Dahabreh$^{2,3,5}$}\\
\footnotesize{$^{1}$ Pattern Recognition Laboratory, Delft University of Technology, Delft, the Netherlands} \\
\footnotesize{$^{2}$ CAUSALab, Harvard T.H. Chan School of Public Health, Boston, Massachusetts, U.S.A}\\
\footnotesize{$^{3}$ Department of Epidemiology, Harvard T.H. Chan School of Public Health, Boston, Massachusetts, U.S.A}\\
\footnotesize{$^{4}$ Department of Computer Science, ETH Zürich, Zürich, Switzerland}\\
\footnotesize{$^{5}$ Department of Biostatistics, Harvard T.H. Chan School of Public Health, Boston, Massachusetts, U.S.A}\\
\footnotesize{$^{\dagger}$Equal contribution; *guanbo.wang@dartmouth.edu; r.k.a.karlsson@tudelft.nl}}\date{}
  \maketitle
} \fi

% \bigskip
\begin{abstract}
One approach for increasing the efficiency of randomized trials is the use of ``external controls'' -- individuals who received the control treatment studied in the trial during routine practice or in prior experimental studies. Existing external control methods, however, can be biased if the populations underlying the trial and the external control data are not exchangeable. Here, we characterize a randomization-aware class of treatment effect estimators in the population underlying the trial that remain consistent and asymptotically normal when using external control data, even when exchangeability does not hold. We consider two members of this class of estimators: the well-known augmented inverse probability weighting trial-only estimator, which is the efficient estimator when only trial data are used; and a potentially more efficient member of the class when exchangeability holds and external control data are available, which we refer to as the optimized randomization-aware estimator. To achieve robust integration of external control data in trial analyses, we then propose a combined estimator based on the efficient trial-only estimator and the optimized randomization-aware estimator. We show that the combined estimator is consistent and no less efficient than the most efficient of the two component estimators, whether the exchangeability assumption holds or not. We examine the estimators' performance in simulations and we illustrate their use with data from two trials of paliperidone extended-release for schizophrenia.
\end{abstract}

\noindent%
\vfill

\newpage
\spacingset{1.45} % DON'T change the spacing!

% %%%%%%%%%%%%%%%%%%%%%%%%%%%%%%%%%%%%%%%%%%%%%%%%%%%%%%%%%%%%%%%%%%%%%

% \title[Robust integration of external control data in trials]{Robust integration of external control data in randomized trials}

% \author{Rickard Karlsson$^{1,\dagger*}$\emailx{r.k.a.karlsson@tudelft.nl; g.wang@hsph.harvard.edu}, 
% Guanbo Wang$^{2,3\dagger*}$, Jesse H. Krijthe$^{1}$ and 
% Issa J. Dahabreh$^{2,3,4}$\\
% $^{\dagger}$Equal contribution \\
% $^{1}$ Pattern Recognition Laboratory, Delft University of Technology, Delft, the Netherlands \\
% $^{2}$ CAUSALab, Harvard T.H. Chan School of Public Health, Boston, Massachusetts, U.S.A\\
% $^{3}$ Department of Epidemiology, Harvard T.H. Chan School of Public Health, Boston, Massachusetts, U.S.A\\
% $^{4}$ Department of Biostatistics, Harvard T.H. Chan School of Public Health, Boston, Massachusetts, U.S.A}

% \begin{document}

% %  This will produce the submission and review information that appears
% %  right after the reference section.  Of course, it will be unknown when
% %  you submit your paper, so you can either leave this out or put in 
% %  sample dates (these will have no effect on the fate of your paper in the
% %  review process!)

% \date{{\it Received June} 2024. {\it Revised XXX} XXX.  {\it
% Accepted XXX} XXX.}

% \pagerange{\pageref{firstpage}--\pageref{lastpage}} 
% \volume{XXX}
% \pubyear{YYY}
% \artmonth{ZZZ}

% \doi{XXX}

\label{firstpage}

%  put the summary for your paper here

%  As usual, the \maketitle command creates the title and author/affiliations
%  display 

\maketitle

%%%%%%%%%%%%%%%%%%%%%%%%%%%%%%%%%%%%%%
\section{Introduction}
%%%%%%%%%%%%%%%%%%%%%%%%%%%%%%%%%%%%%%

Randomized trials are the preferred approach for assessing treatment effectiveness. Trials, however, can be costly and time-consuming, and often have small sample sizes and imprecise results. One approach for improving the efficiency of trials involves augmenting them with data from \textit{external} or \textit{historical controls}~\citep{pocock1976combination, jahanshahi2021use} -- individuals who received the control treatment as part of routine care or prior clinical investigations.

For example, the challenges with small sample sizes in trials are particularly pronounced in studies of schizophrenia -- a chronic, severe, and highly disabling condition, often leading to significant social and occupational impairment~\citep{schizophernia2023nimh}. Difficulties with recruiting participants and monitoring their highly variable symptoms -- even in clinical trials -- result in sparse data in schizophrenia studies~\citep{deckler2022challenges}, leading to imprecise estimates of treatment effects. Incorporating external control data is a promising approach for enhancing the efficiency of such clinical trials, as in our motivating examples, presented in Section \ref{sec:Application}.

Augmenting trials with external controls is related to the problem of transporting causal inferences from a trial to a target population \citep{dahabreh2020extending} because the former essentially ``reverse the flow of information'' compared with the latter \citep{ung2024combining}: instead of using information from the trial to learn about a target population, external control methods use information from an external population to improve inference in the trial. Consequently, trial analyses that use external control data often assume exchangeability conditions similar to those needed for transportability analyses \citep{van2018including, ventz2022design, schuler2022increasing, li2023improving, wang2024evaluating, van2024adaptive}.  However, when these exchangeability conditions do not hold, external control methods can introduce significant bias in the estimation of treatment effects. A natural, though imperfect, approach to address this challenge involves conducting a statistical test to assess whether the trial and external control populations are compatible for pooling \citep{yang2023elastic}. Unfortunately, these tests have low statistical power, particularly when the trial sample size is small, which is precisely when using external control data would be most appealing. False negative results may result in substantial bias. The related approach of dynamically selecting valid external controls in a data-driven manner~\citep{viele2014use, gao2023integrating}, is subject to the same risk of bias.

Here, we describe a ``randomization-aware'' class of estimators that can incorporate external control data and remain consistent and asymptotically normal, even when the external control population is not exchangeable with the trial population. We use optimization methods to identify a potentially more efficient member of this class when exchangeability holds; we refer to this member as the optimized randomization-aware estimator. Moreover, we propose a combined estimator that, asymptotically, is no less efficient than the most efficient trial-only estimator and the optimized randomization-aware estimator, whether the exchangeability assumption holds or not. In simulation studies, we verify that our estimators have good finite-sample performance and are competitive with existing, less robust alternatives. Last, we apply the methods to data from two trials of paliperidone extended-release for~schizophrenia.

\section{Study design, data structure, and causal estimands}
\label{sec:study_design_data_estimands}

\textbf{Study design and data structure:} We assume that the trial and external control data are independently obtained simple random samples from different underlying populations, with unknown and possibly different sampling fractions. The trial and external control data are appended to form a composite dataset. In prior work, this sampling scheme has been referred to as a non-nested trial design because the proportions of trial participants and external controls in the composite dataset do not necessarily reflect the relative size of the underlying populations~\citep{dahabreh2021study}; see also Supplementary Material~\ref{app:study_design}.

\textbf{Simplifying assumptions:} To focus on issues related to the integration of external controls, we make several simplifying assumptions: we consider only binary treatments and we assume complete adherence to treatment, no missing data, and no loss to follow-up. Standard methods for addressing these complications can be combined with the approaches we focus on. 

\textbf{Notation:} We use italic capital letters for random variables and lowercase letters for specific values. We denote densities of random variables by $f(\cdot)$. Let $X\in\mathcal{X}$ denote baseline (pre-randomization and pre-treatment) covariates where $\mathcal{X}$ is the support over all possible covariate patterns, $S$ the binary indicator for the study source ($S=1$ for trial participants; $S = 0$ for the external controls), $A$ the treatment strategy ($A=1$ for the experimental treatment; $A=0$ for the control treatment), and $Y$ the binary, continuous, or count outcome measured at the end of the study.

\textbf{Sampling model:} We model the data on observation $i$ with $S_i = s$ as independent and identically distributed, conditional on study source, realizations of the random tuple $O_i=(X_i, S_i = s, A_i, Y_i)$, for $i=1, \ldots, n_s$, where $n_{s}$ denotes the number of observations from source $S = s$. We define $n = n_{1} + n_{0}$ as the sample size of the composite dataset. In the trial, treatment $A$ is randomly assigned. In the population underlying the external control data, the only treatment in use may be the control treatment, in which case $\{S_i = 0\} \implies \{A_i = 0\}$, or  treatment may be more variable, including the experimental and control treatments evaluated in the trial, as well as other treatments not examined in the trial. To simplify exposition, we mainly address the case of uniform use of the control treatment in the population underlying the external control data; we illustrate this data structure in Table~\ref{tab:data_structure}. Nevertheless, with small modifications, the methods we propose can also be applied when there exists variation in treatment in the population underlying the external data. Regardless of whether there is treatment variation, in many applied settings, the number of external controls, $n_0$, is much larger than the number of trial participants $n_1$. 
With increasing total sample size, we assume that the ratios of the sample sizes of the trial and external control data over the total sample size converge to positive constants, that is, as $n\rightarrow\infty$, $n_s/n\rightarrow q_s> 0$.

\begin{table}[ht]
    \centering
    \caption{The data contains information on baseline covariates $X$, an indicator for which population the observation belongs to (trial $\{S=1\}$ or external $\{S=0\}$ population), treatment $A$, and outcome $Y$. We have $n_1$ observations of the trial population and $n_0$ from the population underlying the external data, and $S=0$ implies $A=0$. }
    \label{tab:data_structure}
    \includegraphics[width=0.5\textwidth]{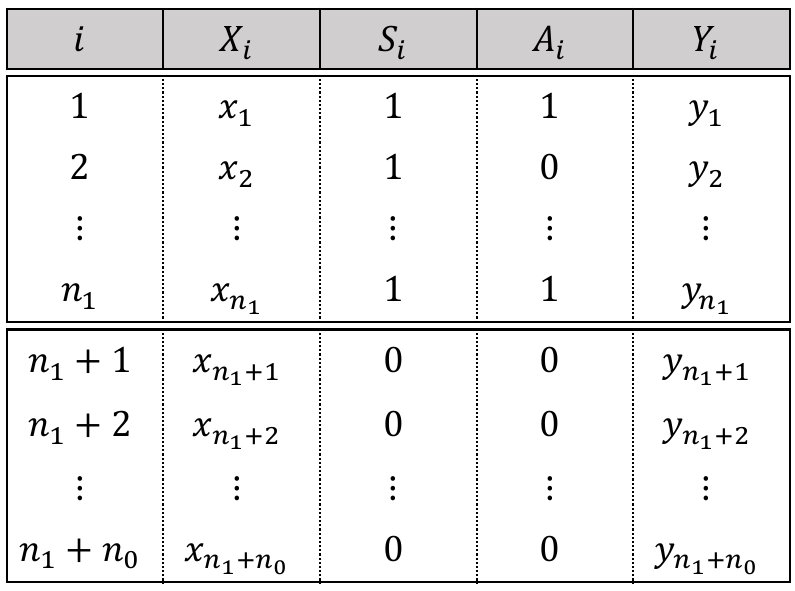}
\end{table}

\textbf{Causal estimands:} To define causal quantities, we use potential (counterfactual) outcomes \citep{rubin1974estimating, robins2000d}. Specifically, for the $i$th individual and for $a\in\{0,1\}$, the potential outcome $Y_i^a$ denotes the outcome under intervention to set treatment $A$ to $a$, possibly contrary to fact. Our goal is to estimate the average treatment effect in the population underlying the trial, $\E[Y^1 - Y^0 | S = 1] = \E[Y^1 | S = 1] - \E[Y^0 | S = 1]$, and its constituent potential outcome means, $\E[Y^a | S = 1]$, $a = 0,1$.

%%%%%%%%%%%%%%%%%%%%%%%%%%%%%%%%%%%%%
\section{Identification and estimation in the trial}
\label{sec:estimation_in_trial}
%%%%%%%%%%%%%%%%%%%%%%%%%%%%%%%%%%%%%

\subsection{Identification in the trial}

\textbf{Identifiability conditions:} The following conditions suffice to identify potential outcome means and average treatment effect in the population underlying the trial: 

% \begin{thmcond}\label{asmp:consistency}
\emph{Condition 1:} For every individual $i$ and each treatment $a \in \{0,1\}$, if $A_i = a,$ then $Y^a_i = Y_i$.
% \end{thmcond}

% \begin{thmcond}\label{asmp:exchangeability}
\emph{Condition 2:} For each treatment $a \in \mathcal \{0,1\}$, $Y^a \indep A | (X, S=1)$. 
% \end{thmcond}

% \begin{thmcond}\label{asmp:positivity}
\emph{Condition 3:} For each treatment $a \in \mathcal \{0,1\}$, if $f(x, S=1) \neq 0$, then $\Pr[A = a | X = x, S = 1] > 0$.
% \end{thmcond}

Condition 1 holds when the intervention is well-defined, such that there is no ``hidden'' or outcome-relevant treatment variation, and there is no interference. This condition is assumed on the basis of substantive knowledge, but aspects of experimental design (e.g., detailed treatment protocol) can increase plausibility. Implicit in our notation is an assumption that data source-specific effects (e.g., trial engagement effects \citep{dahabreh2019generalizing}) are absent. Condition 2 is an assumption of no unmeasured confounding in the trial. This assumption is supported by study design in the context of a marginally or conditionally randomized trial. Condition 3, also supported by design, ensures that every covariate pattern in the trial population has a non-zero probability of receiving each treatment.

\textbf{Identification:} Under conditions 1 through 3, the trial data alone can be used to identify the potential outcome mean under intervention to set treatment $A$ to $a$ in the trial population, $\E[Y^a|S=1]$ , with $\psi_{a} \equiv \E \big [\E[ Y | X, S =1 , A = a ]\big | S = 1 \big ] = \dfrac{1}{\Pr[S = 1]} \E \left[ \dfrac{\mathbf{1}(S = 1, A = a)Y}{\Pr[A=a|X, S=1]} \right].$ Furthermore, the average treatment effect in the population underlying the trial, $\E[Y^1-Y^0|S=1]$, is identified with $\tau=\psi_1-\psi_0.$

\subsection{Estimation using trial data alone}
To estimate $\psi_{a}$ , we can use an outcome regression estimator $\left(\sum_{i=1}^{n} S_i \right)^{-1}\sum_{i=1}^{n} S_i \widehat g_{a}(X_i)$, where $\widehat{g}_{a}(X)$ is an estimator for $g_a(X) = \E[Y|X, S=1, A=a]$. When the model for $g_{a}(X_i)$ is correctly specified, the outcome regression estimator is consistent. However, correct specification of the outcome model is challenging. Alternatively, we can estimate $\psi_{a}$ using the inverse probability weighting estimator $\left(\sum_{i=1}^{n} S_i \right)^{-1}\sum_{i=1}^{n} S_i \mathbf{1}(A_i=a)Y_i/ e_a(X_i)$, where $ e_a(X)= \Pr[A=a|X, S=1]$ is the propensity score in the trial \citep{rosenbaum1983central}. This weighting estimator is consistent because the propensity score in the trial is known by design. 

An estimator that combines the advantages of the outcome regression and weighting estimators, is the augmented inverse probability weighting estimator,
$$
 \widehat \phi_{a} = \left(\sum_{i=1}^{n} S_i \right)^{-1}\sum_{i=1}^{n } S_i\left[\dfrac{\mathbf{1}(A_i=a)}{ e_{a}(X_i)}\left(Y_i-\widehat g_{a}(X_i)\right) + \widehat g_{a}(X_i) \right].
$$
This estimator is asymptotically normal and \emph{robust}, in the sense that it remains consistent even if the outcome regression model for $g_{a}(X)$ is misspecified, because the propensity score $e_{a}(X)$ is known. When the model for $g_{a}(X)$ is correctly specified, the estimator achieves the semiparametric variance bound using trial data alone \citep{robins1994estimation, robins1995semiparametric}. A natural estimator for the average treatment effect in the trial $\tau$ is $\widehat \tau(\widehat g)=\widehat \phi_{1}-\widehat \phi_{0}$; here, we index the estimator by $\widehat g$ to emphasize that it depends on $\widehat g_a(X)$, $a=0, 1$. Due to the linearity of the influence functions, $\widehat \tau(\widehat g)$ is consistent and asymptotically normal as well. We refer to $\widehat \tau(\widehat g)$ as the \textit{efficient trial-only estimator} of the treatment effect. In the next section, we consider strategies for incorporating external control~data.

\section{Using external controls under conditional exchangeability}\label{sec:using_exchangeability_conditions}

We now present additional conditions that are often invoked when using external control data. In view of the above results, these additional conditions are not necessary for identification of the causal estimands; rather, they are invoked in the hopes of improving efficiency.

\subsection{Identification using the external control data}

In prior work (e.g.,~\cite{li2023improving, valancius2023causal}), some version of the following two conditions has been invoked to allow the use of external control data:

\emph{Condition 4:} $Y^{0}\indep S \mid X$.

\emph{Condition 5:} $\{S=0\}\implies\{A=0\}$.

Alternatively, Condition 5 can be replaced with the following independence condition: 

\emph{Condition 5':} $Y^0\indep A|(X, S=0)$.

Condition 4 is a condition of exchangeability in distribution that allows the external control data to contribute to the analysis of the trial. In our setting, it can be replaced by the weaker condition of exchangeability in expectation, $\E[Y^0 | X = x, S = 0] = \E[Y^0 | X = x, S = 1]$, for each $x$ with $f(x,S=1) \neq 0$. Condition 5 states that all individuals in the population underlying the external control data receive the control treatment in the trial. As noted, this condition can be replaced with Condition 5', a partial (only under $a=0$) no-confounding assumption; this assumption may be plausible even in the presence of treatment variation in the population underlying the external data. We reiterate that conditions 4, and 5 or 5', are strong assumptions, not supported by study design, and of uncertain plausibility. However, when they hold, using external control data offers the potential for improved efficiency.

\subsection{Estimation assuming exchangeability of populations}

Estimators for integrating external control data under exchangeability of the trial and external population can be categorized into two types. The first category of methods, such as those in \citet{li2023improving} and \citet{valancius2023causal}, fully pool trial and external control data under conditions 4 and 5. While these methods offer efficiency gains when these conditions hold, they become biased if the conditions are violated. To address this, one can use the fact that conditions 1–3, along with 4 and 5, impose testable restrictions on the observed data (an example of overidentification). This allows for a test-then-pool approach -- an example of pre-test estimation -- that statistically tests the restrictions, and uses the trial-only estimator when the test rejects or a pooling estimator when the test does not reject. This approach performs poorly when the test has low statistical power. A more detailed discussion on these methods is provided in Supplementary Material~\ref{app:estimation_under_exchangeability}.

The second category of estimators borrows a subset of the external control data in a data-driven manner, ensuring compatibility with the exchangeability conditions~\citep{viele2014use, gao2023integrating}. Such borrowing might reduce bias when exchangeability does not hold, but it still carries more risk than necessary: mistakenly pooling trial data with incompatible external control data can introduce bias that could have been avoided by limiting the analysis to the trial data.

%%%%%%%%%%%%%%%%%%%%%%%%%%%%%%%%%%%%%
\section{A novel estimation approach using external controls}
%%%%%%%%%%%%%%%%%%%%%%%%%%%%%%%%%%%%%

Our aim is to develop a \textit{consistent} estimator that leverages the external control data to improve efficiency in the trial when exchangeability between sources holds, but does not rely on exchangeability assumptions for consistency. In outline, our strategy is as follows: First, we examine a class of randomization-aware estimators for $\tau$ that are consistent when conditions 1 through 3 hold; the efficient trial-only estimator is a member of this class. Next, we identify a member within the class that has improved efficiency when conditions 4 and 5 hold. Last, we introduce a combined estimator based on the efficient trial-only estimator and the optimized randomization-aware estimator, and show that it only requires conditions 1 through 3 for consistency, and is no less efficient than the most efficient of the efficient trial-only  estimator and the optimized randomization-aware estimator.

%%%%%%%%%%%%%%%%%%%%%%%%%%%%%%%%%%%%%
\subsection{A class of consistent estimators}\label{sec:method_1}
%%%%%%%%%%%%%%%%%%%%%%%%%%%%%%%%%%%%%

We consider the following class of estimators indexed by a function $h:\mathcal X \to \mathbb R$, 
$$
\widehat \psi_{a} (h) =  \left(\sum_{i=1}^{n} S_i \right)^{-1}\sum_{i=1}^{n } S_i\left[\frac{\mathbf{1}(A_i=a)}{e_a(X_i)}\{Y_i- h(X_i)\} + h(X_i) \right],
$$
which can be viewed as the solution to the estimating equation $\sum_{i=1}^{n } S_i\big[\frac{\mathbf{1}(A_i=a)}{e_a(X_i)}\{Y_i- h(X_i)\} +$ $h(X_i)- \widehat \psi_{a} (h) \big]=0$.
Different choices of the function $h$ correspond to different estimators $\widehat \psi_{a} (h)$, each with different properties. For instance, when $h \equiv 0$, the resulting estimator $\widehat \psi_{a} (0)$ is the inverse probability weighting estimator. Similarly, when $h \equiv \widehat g_a$, the resulting estimator $\widehat \psi_{a} (\widehat g_a)$ is the augmented inverse probability weighting estimator $\widehat \phi_{a}$. 

We prove in Supplementary Material~\ref{app:robust} that the estimator $\widehat \psi_{a} (h)$ is robust (i.e., consistent regardless of the specification of $h$). We refer to this class of estimators as \textit{randomization-aware} because its members exploit randomization and knowledge of the probability of treatment to ensure consistency. Instances of specific randomization-aware estimators have appeared in previous work on using external data sources (e.g.,~\cite{gagnon2023precise}).

%It is sometimes preferred to estimate the probability of treatment even when it is known (see, e.g., \citep{williamson2014variance}). However, for randomization-aware estimators in specific, using the true propensity score $e_a(X)$ will be preferred to construct an estimator integrates external control data and remains provably unbiased even if the underlying external control population is incompatible with the underlying trial population.

%Note that even though the probability of treatment is known by design in our setting, it is often preferable to estimate it to improve efficiency (see, e.g., \citep{williamson2014variance}). In trial analyses, it is reasonable to do so using a parametric model (e.g., a logistic regression model) that can be estimated at $\sqrt n$-rate of convergence. 

To borrow information from the external control population, we attempt to ``learn'' $h(X)$ using both the trial and external control data. Throughout this paper, we assume that $h(X)$ satisfies certain regularity conditions; namely, that it is continuous, differentiable, smooth, and has finite expectation and variance at all points for $X=x$. Informally, the purpose of $h(X)$ in our estimators is to capture the relationship between the outcome and baseline covariates; flexible regression models such as splines and kernel smoothing methods can approximate the relationship and satisfy the regularity conditions.

Consider the choice $h \equiv  h_{\text{fix}}$, when the estimator $\widehat \psi_{a}( h_{\text{fix}})$ depends on the true propensity score and some fixed function $h_{\text{fix}}$ satisfying the above-mentioned regularity conditions. We note the following properties of $\widehat \psi_{a}( h_{\text{fix}})$ (see Supplementary Material \ref{app:lemma:fix} for the proof).
\begin{thmlem}\label{lemma:fix}
Suppose $Y$ has finite mean and variance, then the estimator $\widehat \psi_{a}(h_{\text{fix}})$ is consistent, asymptotically normal with asymptotic variance
$\Pr[S=1]^{-1}[\mathcal{C}_a+f_a(h_{\text{fix}}(X), Y)]$, where
$\mathcal{C}_a=\text{Var}[Y^a|S=1]$, and 
\begin{align}\label{eq:AV_fix}
    f_a(h_{\text{fix}}(X), Y)=\E\left[\Pr[A=a|S=1]\frac{1-e_{a}(X)}{e_{a}^2(X)}\{Y-h_{\text{fix}}(X)\}^2 \middle\vert S=1, A=a\right].
\end{align}
\end{thmlem}

In other words, $\widehat \psi_{a}(h_{\text{fix}})$ is asymptotically normal with asymptotic variance whose only component that involves $h_{\text{fix}}$ is $f_a(h_{\text{fix}}(X), Y)$. We will exploit Lemma \ref{lemma:fix} to develop a procedure for choosing $h$ that attempts to improve the efficiency of our estimators.

%%%%%%%%%%%%%%%%%%%%%%%%%%%%%%%%%%%%%
\subsection{An optimized randomization-aware estimator}\label{sec:method_2}
%%%%%%%%%%%%%%%%%%%%%%%%%%%%%%%%%%%%%

Consider the class of randomization-aware estimators  $\widehat \psi_{a}(h)$. The results in the previous sub-section suggest that we can ``learn'' an $h(X)$ function that results in a more efficient randomization-aware estimator. Specifically, we propose to use the trial and external control data together to find $h(X)$ that minimizes the asymptotic variance of $\widehat \psi_{0}(h_{\text{fix}})$, which is the same as minimizing the term $f_0(h_{\text{fix}}(X), Y)$ in \eqref{eq:AV_fix}. Our approach is similar to the approach of \citet{cao2009improving} for the problem of estimating an expectation of an outcome with missing data in a single data source setting. 

In the results presented above, however, $f_0(h_{\text{fix}}(X), Y)$ is written as an expectation conditional on $S=1$. To incorporate external control data in the optimization, we use conditions 4 and 5 to rewrite $f_0(h_{\text{fix}}(X), Y)$ as an expectation over both trial and external control data (see Supplementary Material \ref{app:lemma:reweite} for proof).

\begin{thmlem}\label{lemma:reweite}
    Under conditions 1 through 5, we have that
\begin{align}\label{eq:rewrite}
    f_{0}(h_{\text{fix}}(X), Y) = \E\left[\frac{\Pr[S=1\mid X, A=0]}{\Pr[S=1\mid A=0]}\frac{\Pr[A=0|S=1]e_1(X)}{e_0^2(X)} \{Y-h_{\text{fix}}(X)\}^2\middle\vert A=0 \right]. 
\end{align}
\end{thmlem}

This lemma provides a way for choosing $h(X)$ using external control data such that the asymptotic variance of $\widehat \psi_{0}( h_{\text{fix}})$ is minimized. Ignoring normalizing constants, we define $h^{*}(X)$ as the minimizer of $R(\widetilde h)=\E\left[\eta_0(X) e_1(X)/e_0^2(X) \{Y-\widetilde h(X)\}^2\middle\vert A=0 \right]$ within a model class $\widetilde{h} \in \mathcal{H}$, where $\eta_0(X) = \Pr[S=1\mid X, A=0]$ is the probability of participation in the trial, conditional on covariates, among individuals receiving treatment $A = 0$. We can estimate $h^{*}(X)$ by finding the minimizer of the sample analog of $R(\widetilde h)$ with the estimated $\eta_0(X)$. We let $\mathcal{H}$ be a class of parametric models. Therefore, $\widetilde h(X)$ can be denoted by $h(X; \widetilde \gamma)$ and $h^{*}(X)$ can be denoted by $h(X; \gamma^{*})$. Thus, estimating $h^{*}(X)$ is the same as estimating  $\gamma^{*}$, which can be obtained by $\widehat{\gamma}^*=\argmin_{\widetilde \gamma}\sum_{i=1}^n\widehat R(O_i; \widetilde \gamma)$, where
$
\widehat R(O_i; \widetilde \gamma)= (1-A_i)  \widehat \eta_0(X_i) e_1(X_i)/ e_0^2(X_i) \{Y_i- h(X_i; \widetilde \gamma)\}^2.
$

\subsection{Implementation using M-estimation methods}

We propose to use M-estimation \citep{stefanski2002calculus} to implement the randomization-aware estimator $\widehat \psi_{0}(\widehat h^{*})$ using $\widehat h^{*}(X)$ as the estimated optimized $h(X)$. For a set of smooth finite-dimensional target parameters $\boldsymbol{\theta}$, its M-estimator $\boldsymbol{\widehat \theta}$ is the solution to a stack of equation of the form $\sum_{i=1}^n\boldsymbol{m}(O_i; \boldsymbol{\widetilde \theta})=\boldsymbol{0}$, where $\boldsymbol{\widetilde \theta}$ is the set of parameters with arbitrary values and $\boldsymbol{m}(O_i; \boldsymbol{\widetilde \theta})$ is the stack of estimating functions.

We consider parametric models for $\eta_0(X)$ and $h^{*}(X)$ denoted by $\eta_0(X; \beta)$ and $h(X;\gamma^{*})$, respectively, and furthermore, we denote $q=\Pr[S=1]$. The set of target parameters is $\boldsymbol{\theta}=\{q, \beta, \gamma^{*}, \psi_{0}\}$ and the stack of estimating functions is
\begin{equation}\label{eq:stack_psi}
\boldsymbol{m}(O_i;\boldsymbol{\widetilde \theta})=\begin{pmatrix}
    m_{q}(O_i; \widetilde q),\;\;
    m_{\eta_0}(O_i; \widetilde \beta), \;\;
    m_{h^{*}}(O_i; \widetilde \beta, \widetilde \gamma), \;\;
    m_{\psi_0}(O_i; \widetilde q, \widetilde \gamma, \widetilde \psi_0)
\end{pmatrix}^\top.
\end{equation}
To estimate $q$, we define $m_{q}(O_i; \widetilde q)=S_i - \widetilde q$. When $\eta_0(X)$ is estimated by logistic regression, $m_{\eta_0}(O_i; \widetilde \beta)$ is the logistic regression score equation. For estimating $\gamma^{*}$ and $\psi_0$, we define 
\begin{gather*}
    m_{h^{*}}(O_i; \widetilde \beta, \widetilde \gamma)=\frac{\partial}{\partial \widetilde \gamma}\widehat R(O_i; \widetilde \beta, \widetilde \gamma)= (1-A_i)  \eta_0(X_i; \widetilde \beta) \frac{e_1(X_i)}{\{1-e_1(X_i)\}^2} \{Y_i-  h(X_i; \widetilde \gamma)\}\frac{\partial}{\partial \widetilde \gamma}  h(X_i; \widetilde \gamma), \\
    m_{\psi_0}(O_i;  \widetilde q, \widetilde \gamma, \widetilde \psi_0)=\dfrac{S_i}{\widetilde q}\left[\frac{\mathbf{1}(A_i=0)}{1-e_{1}(X_i)}\{Y_i- h(X_i; \widetilde \gamma)\} +  h(X_i; \widetilde \gamma) - \widetilde \psi_0\right].
\end{gather*}

We obtain a consistent estimator of $\widehat \psi_{0}( \widehat h^{*})$ by jointly solving the stack of estimating functions, that is, letting $\boldsymbol{\widehat \theta}$ be the solution to $\sum_{i=1}^n \boldsymbol{m}(O_i; \boldsymbol{\widetilde \theta}) = \boldsymbol{0}$. The following theorem summarizes the properties of $\widehat \psi_{0}(\widehat h^{*})$ obtained from solving this optimization task (see Supplementary Material~\ref{app:them:efficient} for the proof). 
\begin{thmthm}\label{them:efficient}
Suppose $\boldsymbol{m}(O_i; \boldsymbol{\widetilde \theta})$ satisfies the regularity conditions described in \ref{app:regularity}; then, the M-estimator $\widehat \psi_{0}( \widehat h^{*})$ is unbiased, consistent, and asymptotically normal. In addition, if conditions 1 through 5 hold and the model for estimating $\eta_0(X)$ can be correctly specified, $\widehat \psi_{0}(\widehat h^{*})$ minimizes the asymptotic variance of $\widehat \psi_{0}(h)$ over $\mathcal{H}$. 
\end{thmthm}

Because $h^{*}(X)$ is not estimated using only trial data ($S=1$), Theorem \ref{them:efficient} provides a practical solution for estimating $\psi_{0}$ when incorporating external control data. Furthermore, we can obtain an estimator for the average treatment effect in the trial population $\tau$ as
\begin{equation}
\label{eq:eff_randomization_aware}
\widehat \tau(\widehat h^{*})=\widehat \psi_1( \widehat g_1)-\widehat \psi_{0}(\widehat h^{*})~.
\end{equation}
Because both of $\widehat \psi_1( \widehat g_1)$ and $\widehat \psi_{0}( \widehat h^{*})$ are consistent estimators, $\widehat \tau(\widehat h^{*})$ is also a consistent estimator. To construct asymptotically normal estimators, we again propose to obtain $\{\widehat \psi_1( \widehat g_1), \widehat \psi_{0}(\widehat h^{*})\}$ via joint M-estimation (see details in Supplementary Material \ref{app:MforTau}), so that $\widehat \psi_1(\widehat g_1)$ and $\widehat \psi_{0}(\widehat h^{*})$ are asymptotically bivariate normal \citep{stefanski2002calculus}. Because a linear combination of two bivariate normally distributed random variables is normally distributed, $\widehat \tau(\widehat h^{*})$ is asymptotically normal.

As a final remark, the consistency of $\widehat \tau(\widehat h^{*})$ does not rely on conditions 4 and 5; however, the efficiency improvement that this estimator hopes to offer depends on these conditions, as well as the specification for the model for $\Pr[S=1|X, A = 0]$. If conditions 4 and 5 do not hold, or if the model for $\Pr[S=1|X, A = 0]$ is misspecified, then $\widehat \tau(\widehat h^{*})$ may be less efficient than the efficient trial-only estimator, $\widehat \tau(\widehat g)$. To further relax the dependence of the estimator's efficiency on these additional conditions that are not justified by randomization, in the next section we develop a new estimator that is asymptotically guaranteed to not perform worse than the efficient trial-only estimator.

%%%%%%%%%%%%%%%%%%%%%%%%%%%%%%%%%%%%%
\subsection{Combined estimator}\label{sec:method_3}
%%%%%%%%%%%%%%%%%%%%%%%%%%%%%%%%%%%%%

We have two different consistent estimators: the efficient trial-only estimator $\widehat \tau(\widehat g)=\widehat \phi_1-\widehat \phi_0$; and the optimized randomization-aware estimator $\widehat \tau(\widehat h^{*})=\widehat \psi_1(\widehat g_1) - \widehat \psi_{0}( \widehat h^{*})$ that incorporates external control data. When conditions 4 and 5 also hold, and necessary statistical models are correctly specified, $\widehat \tau(\widehat h^{*})$ is expected to be more efficient (in finite-sample) than $\widehat \tau(\widehat g)$ because it uses more observations to model the outcome conditional on covariates, under the control treatment. However, when conditions 4 and 5 do not hold, the relative efficiency of $\widehat \tau(\widehat h^{*})$ and $\widehat \tau(\widehat g)$ depends on many factors, including the sample sizes of the trial and external controls and the extent to which these conditions are violated. To avoid choosing between $\widehat \tau(\widehat h^{*})$ and $\widehat \tau(\widehat g)$, we consider combining these two estimators, in a way that may provide further efficiency gains~\citep{graybill1959combining}. Specifically, we propose the combined~estimator
\begin{equation*}
\label{eq:combined_estimator}
\widehat \tau(\lambda)=\lambda \widehat \tau(\widehat h^{*})+(1-\lambda)\widehat \tau(\widehat g),  \;\;\forall \lambda\in\mathbb{R}.
\end{equation*}
If $\lambda=0$, then $\widehat \tau(\lambda)$ degenerates to the efficient trial-only estimator $\widehat \tau(\widehat g)$; in all other cases,  $\widehat \tau(\lambda)$ incorporates information from the external control data. We note some important properties of $\widehat \tau(\lambda)$ that hold for all $\lambda\in\mathbb{R}$ (see Supplementary Material~\ref{app:lem:combined_consistency} for the proof).
\begin{thmlem}\label{lem:combined_consistency}
    If $\{\widehat \tau(\widehat g), \widehat \tau(\widehat h^{*})\}$ is obtained via joint M-estimation, $\widehat \tau(\lambda)$ is unbiased, consistent and asymptotically normal for all $\lambda\in\mathbb{R}$.
\end{thmlem}
With our combined estimator being consistent, we propose to select $\lambda$ such that its asymptotic variance, denoted by $\samplevar_{\lambda}$, is minimized. This equates to solving $\lambda^{*}=\argmin_{\lambda}\samplevar_{\lambda}$. Denote the asymptotic variance of the estimators $\widehat \tau(\widehat g)$ and $\widehat \tau(\widehat h^{*})$ by $\samplevar_{g}$ and $\samplevar_{h^{*}}$, and the asymptotic covariance by $\samplestd_{g, h^{*}}$. By writing $\samplevar_{\lambda}=\lambda^2\samplevar_{h^{*}}+(1-\lambda)^2\samplevar_{g}+2\lambda(1-\lambda)\samplestd_{g,h^{*}}$, we see that $\samplevar_{\lambda}$ is a quadratic function of $\lambda$ which has closed-form expressions for the optimal $\lambda^{*}$ and the corresponding variance $\samplevar_{\lambda^{*}}$. 
$$
 \lambda^{*}=\dfrac{\samplevar_{g}- \samplestd_{g, h^{*}}
}{
 \samplevar_{g}+ \samplevar_{h^{*}}-2 \samplestd_{g, h^{*}}
},
\quad \text{and} \quad
\samplevar_{\lambda^{*}}=\dfrac{\samplevar_{g}\samplevar_{h^{*}}- \samplevar_{g, h^{*}}
}{
 \samplevar_{g}+ \samplevar_{h^{*}}-2 \samplestd_{g, h^{*}}
}.
$$
%The optimal  $\lambda^{*}$ is well-defined as long as the denominator $\samplevar_{g}+ \samplevar_{h^{*}}-2 \samplestd_{g, h^{*}} = \samplevar_{g-h^{*}} \neq 0$
In Supplementary Material~\ref{app:existlambda}, we show that $\lambda^*$ is well-defined unless $\widehat g$ and $\widehat h^*$ converge to the same asymptotic limit. Since we expect both of these models to be misspecified, in what follows, we assume that $\widehat g $ and $\widehat h^*$ converge to different limits, and $\lambda^*$ is well-defined. 
In practice, $\lambda^{*}$ is unknown because the asymptotic variances and covariance of $\widehat \tau(\widehat g)$ and $\widehat \tau(\widehat h^{*})$ are unknown. We can estimate $(\samplevar_{g}, \samplevar_{h^{*}}, \samplestd_{g, h^{*}})$ by the sandwich estimators as usually done in M-estimation, denoted by  $(\hatsamplevar_{g}, \hatsamplevar_{h^{*}},\widehat\samplestd_{g, h^{*}})$. Then we can estimate $\lambda^*$ and $\samplevar_{\lambda^{*}}$ using plug-in estimators, denoted by $\widehat\lambda^{*}$ and $\widehat \samplevar_{\widehat \lambda^{*}}$.
% as follows,
% \begin{equation} \label{eq:optimal_lambda}
%   \widehat\lambda^{*}=\dfrac{ \widehat\sigma^2_{g}-  \widehat\sigma_{g, h^{*}}
% }{
%   \widehat\sigma^2_{g}+  \widehat\sigma^2_{h^{*}}-2  \widehat\sigma_{g, h^{*}}
% },
% \quad \text{and} \quad
%  \widehat\sigma^{2}_{\widehat \lambda^{*}}=\dfrac{ \widehat\sigma^2_{g} \widehat\sigma^2_{h^{*}}-  \widehat\sigma^2_{g, h^{*}}
% }{
%   \widehat\sigma^2_{g}+  \widehat\sigma^2_{h^{*}}-2  \widehat\sigma_{g, h^{*}}
% }.
% \end{equation}

Because the sandwich estimators are consistent estimators of the respective asymptotic variances~\citep{stefanski2002calculus}, $\widehat\lambda^{*}$ is a consistent estimator for $\lambda^{*}$. More importantly, we can show that the (feasible) estimator $\widehat\tau (\widehat\lambda^*)$ using the plugin estimator $\widehat\lambda^*$ converges asymptotically to the same distribution as the (infeasible) estimator $\widehat\tau (\lambda^*)$ that requires oracle knowledge of $\lambda^*$ (see Supplementary Material~\ref{app:thm_combine_optimal} for the proof). 
\begin{thmthm}\label{thm_combine_optimal} 
The estimator $\widehat\tau (\widehat\lambda^*)$ is asymptotically equivalent to $\widehat\tau (\lambda^*)$; that is, $\sqrt{n}(\widehat\tau (\widehat\lambda^*) - \tau) \xrightarrow{d} N(0,\samplevar_{\lambda^{*}})$ where $\xrightarrow{d}$ denotes convergence in distribution. In addition, $\widehat\tau (\widehat\lambda^*)$ is no less efficient than $\widehat\tau (\widehat g)$ and $\widehat\tau (\widehat h^*)$; i.e.,   $ \sigma^2_{ \lambda^{*}} \leqslant \text{min}\{ \sigma^2_{g},  \sigma^2_{h^{*}}\}$.
\end{thmthm}
 
In sum, $\widehat \tau(\widehat \lambda^{*})$ is robust in the sense that whether additional conditions hold or not, it is consistent for $tau$. Furthermore, its efficiency is no less than that of its component estimators.

\section{Simulation studies}
We evaluated our estimators' finite-sample performance against existing methods for integrating external control data in clinical trials via simulations. First, we considered a scenario favorable to all methods, where conditions 1-5 hold, all parametric models are correctly specified, and there is no distribution shift in baseline characteristics. Second, we considered a more realistic scenario where conditions 4-5 do not hold, parametric models are misspecified, and there is a distribution shift, such that $f(X\mid S=1) \neq f(X\mid S=0)$. Detailed simulation methods and results are presented in Supplementary Material~\ref{app:simulation_studies}. In both scenarios, our estimators were more efficient than the trial-only estimator and remained nearly unbiased, while alternatives showed significant biases in the more realistic scenario. When comparing trial sizes of 50 or 200, the greatest variance reduction occurred with the smaller trial size; with larger trials, the variance improvements for all methods were less pronounced.

%%%%%%%%%%%%%%%%%%%%%%%%%%%%%%%%%%%%%%%%%%%%%%%
\section{Augmenting a trial of treatments for schizophrenia using external controls}\label{sec:Application}
%%%%%%%%%%%%%%%%%%%%%%%%%%%%%%%%%%%%%%%%%%%%%%%
To illustrate the proposed methods, we used data from two independent placebo-controlled, double-blind trials (NCT00668837 and NCT00077714) that compared the effect of paliperidone ER tablets 6 mg versus placebo on schizophrenia symptoms. We designated one trial \citep{marder2007efficacy} as the index trial (i.e., the trial that we aimed to augment using external data, denoted by $S=1$), and used data from the placebo group in the second trial as external controls (denoted by $S=0$) \citep{davidson2007efficacy}.

Positive and Negative Syndrome Scale (PANSS) total scores are used for rating the severity of schizophrenia symptoms. The outcome of interest in our analyses was the PANSS scores of patients at week 6 after randomization. We included patients assigned to either paliperidone ER or placebo and for whom PANSS scores were available at baseline and at week 6. From the index trial, we included 97 patients who were assigned to paliperidone ER tablets 6 mg, and 91 patients who were assigned to the placebo; the sample size of the external control data was 111. We used patient gender, age, race, and baseline PANSS score as covariates. Their summary statistics are given in Supplementary Material \ref{app:TableOne}. 

Table \ref{tab:ApplicationResults} summarizes results from the unadjusted trial-only, pooling under exchangeability \citep{li2023improving}, test-then-pool, selective borrowing~\citep{gao2023integrating}, and Bayesian dynamic borrowing \citep{viele2014use} estimators, alongside the novel estimators proposed herein.

The point estimate of the unadjusted estimator that uses the trial data only should be trustworthy, which indicates the treatment can lower the PANSS total score by 8 after 6 weeks of treatment initiation. However, the estimates from the pooling and test-then-pool estimators deviate from the -8, meaning that the external data are not compatible with the trial data. On the contrary, the estimates from the three proposed consistent estimators are close enough to -8 with smaller standard errors. The $\widehat \lambda^{*}$ of the combined estimator is 0.064, indicating that the external data and the trial data are not compatible with each other, which also explains the reason why the pooling estimator is inconsistent with the trial-only estimators. The selective borrowing methods produced similar estimates to the combined estimator, while Bayesian dynamic borrowing methods produced biased results.

\begin{table}[ht]{\footnotesize
    \centering
    \caption{Estimates and standard errors of the different estimators.}
    \label{tab:ApplicationResults}
    \begin{tabular}{lcccccccc}\hline  
       &Unadjusted & \shortstack{Pooling}&\shortstack{Test-\\pool}& $\widehat \tau(\widehat g)$  & $\widehat \tau(\widehat h^{*})$ & $\widehat \tau(\widehat \lambda^{*})$ & \shortstack{Selective\\borrowing} &\shortstack{Dynamic\\borrowing}\\\hline
      Estimate &-7.965&-9.562&-9.562  &-7.7726  & -7.532  & -7.714 & -7.711 & -10.038\\
      Standard error & 12.374& 2.460& 2.460 & 2.896  & 2.927 & 2.894 &2.894 & 2.633\\\hline
 \end{tabular}}
\end{table}

We also repeated the analyses by randomly sampling with replacement a fraction ($\sim75\%$, $\sim50\%$, $\sim25\%$) of the available control observations (see Supplementary Materials \ref{app:application_add}). The results showed that $\widehat \tau(\widehat \lambda^{*})$ can generate reasonable estimates with smaller or similar standard errors (relative to its alternatives) as the control group sample size decreases. 

%%%%%%%%%%%%%%%%%%%%%%%%%%%%%%%%%%%%%%%%%%%%%%%%%%%%%%%%%%%%%
\section{Discussion} \label{sec:discussion}
%%%%%%%%%%%%%%%%%%%%%%%%%%%%%%%%%%%%%%%%%%%%%%%%%%%%%%%%%%%%%

We proposed a novel approach for using external control data to improve inference in trials. Like earlier work, we show efficiency gains are possible under exchangeability conditions between the trial and external control populations. However, our optimized randomization-aware estimator explicitly avoids relying on these additional conditions for its consistency, and only uses them in an attempt to improve efficiency. Moreover, by combining the efficient trial-only estimator with our optimized randomization-aware estimator we provide a new estimator that is no less efficient than the most efficient of these two component estimators. The combined estimator may lead to further efficiency gains, but its main attraction is protection from performing worse than the efficient trial-only estimator in large samples.

Throughout, we used parametric M-estimation methods to jointly estimate all nuisance models and the trial-only and optimized randomization-aware estimators. This approach makes the logic of combining information transparent and ensures the joint normality of the two estimators. The majority of trial analyses use simple parametric models; therefore, our approach can be viewed as a relatively natural next step when trial data are to be combined with external control data. One limitation of our approach may be unstable coverage when the sample size is very small. Nevertheless, even in this setting, our estimator had better coverage than common alternatives. Extensions of our approach to use data-adaptive (e.g., machine learning) modeling strategies may further improve performance.

%%%%%%%%%%%%%%%%%%%%%%%%%%%%%%%%%%%%%%%%%%%%%%%
\section*{Acknowledgments}
%%%%%%%%%%%%%%%%%%%%%%%%%%%%%%%%%%%%%%%%%%%%%%%

This work was supported in part by National Library of Medicine (NLM) award 5R01LM013616, National Heart, Lung, and Blood Institute (NHLBI) award R01HL136708, and Patient-Centered Outcomes Research Institute (PCORI) award ME-2021C2-22365. The content is solely the responsibility of the authors and does not necessarily represent the official views of NLM, NHLBI, PCORI, PCORI's Board of Governors or PCORI's Methodology Committee. 

We used data from the Yale University Open Data Access (YODA) registry (\url{https://yoda.yale.edu/}) under project number 2022-5062. YODA has an agreement with Jassen Research \& Development, L.L.C.. The interpretation and reporting of research using these data are solely the responsibility of the authors and do not necessarily represent the views of the Yale University Open Data Access Project or Jassen Research \& Development, L.L.C..
\vspace*{-8pt}

%%%%%%%%%%%%%%%%%%%%%%%%%%%%%%%%%%%%%%%%%%%%%%%
\section*{Data availability}
%%%%%%%%%%%%%%%%%%%%%%%%%%%%%%%%%%%%%%%%%%%%%%%
Code to reproduce our simulations and data analyses is publicly available: \url{https://github.com/RickardKarl/IntegratingExternalControls}.
Data used in this paper to illustrate our findings can be obtained from YODA \url{https://yoda.yale.edu/}, subject to approval.

\section*{References}
Please find the references at the end of this document.

%%%%%%%%%%%%%%%%%%%%%%%%%%%%%%%%%%%%%%%%%%%%%%%
%\section*{Supplementary material}
%%%%%%%%%%%%%%%%%%%%%%%%%%%%%%%%%%%%%%%%%%%%%%%

% \bibliographystyle{apalike}
% \bibliography{references}
\clearpage

\begin{center}
\textbf{Supplementary Materials for \\``Robust integration of external control data in randomized trials''}
\end{center}
\setcounter{page}{1}
\renewcommand{\thepage}{S\arabic{page}}
\begin{appendices}
\numberwithin{equation}{section}
\renewcommand{\theequation}{\thesection.\arabic{equation}}
\numberwithin{table}{section}
\renewcommand{\thetable}{\thesection.\arabic{table}}
\numberwithin{figure}{section}
\renewcommand{\thefigure}{\thesection.\arabic{figure}}
\newenvironment{proof}{\paragraph{Proof:}}{\hfill$\square$}

\section{Proofs}

\subsection{Proof of the robustness property of a special case of $\psi_{a}( h)$}\label{app:robust}
\begin{proof}
We first observe that for any measurable function $h$, $\widehat \psi_{a} ( h) $ converges in probability to the following quantity.
\begin{align*}
    \widehat \psi_{a} ( h) 
    &=\left(\sum_{i=1}^{n} S_i \right)^{-1}\sum_{i=1}^{n } S_i\left[\frac{\mathbf{1}(A_i=a)}{e_a(X_i)}\{Y_i- h(X_i)\} + h(X_i) \right]\\
    &\overset{p}{\to}\dfrac{1}{\Pr[S=1]}\E\left[S\dfrac{\mathbf{1}(A=a)}{e_a(X)}Y\right]+\dfrac{1}{\Pr[S=1]}\E\left[Sh(X)\left\{1-\dfrac{\mathbf{1}(A=a)}{e_a(X)}\right\}\right].
\end{align*}
Further, the first term equals $\E[Y|A=a, S=1]$ and the second term is 0. Therefore, $\widehat \psi_{a} ( h)$ is a consistent estimator for $\psi_a$ regardless of the specification of $h$.
\end{proof}

\subsection{Proof of Lemma \ref{lemma:fix}}\label{app:lemma:fix}
\begin{proof}
% We first prove the estimator is an unbiased estimator under the conditions.
% \begin{align*}
%     \E[\widehat \psi_{a}(h_{\text{fix}})]=&
%     \E\left[\frac{1}{\Pr[S=1]} \frac{1}{n} \sum_{i=1}^{n}S_i\left\{\frac{\mathbf{1}(A_i=a)}{e_{a}(X_i)}\{Y_i-h_{\text{fix}}(X_i)\} + h_{\text{fix}}(X_i)\right\}\right]\\
%     =&\frac{1}{\Pr[S=1]} \frac{1}{n} \sum_{i=1}^{n}\left[\E\left\{\frac{\mathbf{1}(A_i=a)}{e_{a}(X_i)}Y_i S_i\right\}+\E\left\{\frac{\mathbf{1}(A_i=a)-e_{a}(X_i)}{e_{a}(X_i)}h_{\text{fix}}(X_i)S_i \right\}\right]\\
%     =&\frac{1}{\Pr[S=1]} \left(\E\left\{\frac{\mathbf{1}(A_i=a)}{e_{a}(X_i)}Y_i S_i\right\}+\E\left[\E\left\{\frac{\mathbf{1}(A_i=a)-e_{a}(X_i)}{e_{a}(X_i)}h_{\text{fix}}(X_i)\middle\vert X_i, S_i=1 \right\}\right]\right)\\
%     =&\frac{1}{\Pr[S=1]} \E\left\{\frac{\mathbf{1}(A_i=a)}{e_{a}(X_i)}Y_i S_i\right\}\\
%     =&\psi_{a}.
% \end{align*}
We have proved the estimator is consistent in \ref{app:robust}; next, we prove it is asymptotically normal. Define $O_i=(Y_i, S_i, A_i, X_i)$, because $\{O_1, O_2,..., O_n\}$ are independent and identically distributed random variables, by the Central limit theorem, when $Y$ has finite mean and variance, the estimator is asymptotically normal.
% Observing the form of $\widehat \psi_{a}(h_{\text{fix}})$, we can see that it can be estimated by finding the $\widetilde \psi$ that solves the following estimating equation
% $$
%   \sum_{i=1}^{n}S_i\left(\frac{1}{\frac{1}{n}\sum_{i=1}^{n}S_i}S_i\left[\frac{\mathbf{1}(A_i=a)}{ e_{a}(X_i)}\{Y_i- h_{\text{fix}}(X_i)\} +  h_{\text{fix}}(X_i) -\widetilde \psi\right]\right)=0.
% $$
% Therefore, $\widehat \psi_{a}(h_{\text{fix}})$ can be viewed as an M-estimator \citep{huber1964robust, huber1967maximum}, so that if $Y$ has finite mean and variance, $\widehat \psi_{a}(h_{\text{fix}})$ is asymptotically normal \citep{stefanski2002calculus}.
Now we derive its asymptotic variance. 

Denote $\Pr[S=1]$ by $q$, $n^{-1}\sum_{i=1}^{n}I(S_i=1)$ by $\widehat q$, and \(\frac{A}{e_a(X)}(Y-h_{\text{fix}}(X))+h_{\text{fix}}(X)\) by $T(X, Y)$.

We first prove that $\psi_a=\E[(S/q) T(X, Y)]$, which will be used in the later proofs.

We observe that
\begin{align*}
\psi_a & =\E[\E[Y \mid X, A=a, S=1] \mid S=1] \\
& =\E\left[\left.\frac{\Pr[A=a \mid X, S=1] \E[Y \mid X, A=a, S=1]}{e_a(X)} \right\rvert\, S=1\right] \\
& =\E\left[\left.\frac{E[A Y \mid X, S=1]}{e_a(X)} \right\rvert\, S=1\right] \\
& =\E\left[\left.\E\left[\left.\frac{A Y}{e_a(X)} \right\rvert\, X , S=1\right] \right\rvert\, S=1\right] \\
& =\E\left[\left.\frac{A Y}{e_a(X)} \right\rvert\, S=1\right] \\
& =\E\left[\left.\frac{A Y}{e_a(X)}+\frac{e_a(X)-A}{e_a(X)} h_{\text{fix}}(X) \right\rvert\, S=1\right]  \\
& =\E\left[\left.\frac{A}{e_a(X)}(Y-h_{\text{fix}}(X))+h_{\text{fix}}(X) \right\rvert\, S=1\right] \\
& =\E[T(X,Y)|S=1]\\
&=\E[(S/q) T(X, Y)]. 
\end{align*}
The third to the fourth equation holds because  
\[
\E\left[\left.\frac{e_a(X)-A}{e_a(X)} h_{\text{fix}}(X) \right\rvert\, S=1\right]=\E\left[\E\left[\left.\frac{e_a(X)-A}{e_a(X)} h_{\text{fix}}(X)\Big|X, S=1 \right]\right\rvert\, S=1\right]=0.
\]

By definition of $\hat{\psi_a}(h_{\text{fix}}(X))$,
\begin{align*}
\mathbb{P}_{n}\left[\frac{S}{\hat{q}}\left[\frac{A}{e_a(X)}(Y-h_{\text{fix}}(X))+h_{\text{fix}}(X)\right\}-\hat{\psi_a}(h_{\text{fix}}(X))\right]=0,
\end{align*}
which implies 
\begin{align*}
\sum_{i=1}^{n}\left[\frac{S_{i}}{\hat{q}} T(X, Y)-\hat{\psi_a}(h_{\text{fix}}(X))\right]=0.
\end{align*}
By construction 
\begin{align*}
n\{\hat{\psi_a}(h_{\text{fix}}(X))-\psi_a\}=\sum_{i=1}^{n}\left[\frac{S_{i}}{\hat{q}} T(X, Y)-\psi_a\right].
\end{align*}
By Taylor expansion,
\begin{align*}
&\hspace{0.5cm}n\{\hat{\psi_a}(h_{\text{fix}}(X))-\psi_a\}\\
& \approx \sum_{i=1}^{n}\left[\frac{S_{i}}{q} T(X, Y)-\psi_a\right]+\sum_{i=1}^{n} \frac{\partial}{\partial q}\left[\frac{S_{i}}{q} T(X, Y)-\psi_a\right](\hat{q}-q) \\
& =\sum_{i=1}^{n}\left[\frac{S_{i}}{q} T(X, Y)-\psi_a\right]-\sum_{i=1}^{n} \frac{S_{i} T(X, Y)}{q^{2}} \cdot \frac{1}{n} \sum_{i=1}^{n}\left(S_{i}-q\right) \\
& \rightarrow \sum_{i=1}^{n} \frac{S_{i}}{q} T(X, Y)-n \psi_a-\E\left[\frac{ST(X, Y)}{q^{2}}\right] \sum_{i=1}^{n}\left(S_{i}-q\right) \\
& =\sum_{i=1}^{n} \frac{S_{i}}{q} T(X, Y)-n \psi_a-\frac{1}{q} \psi_a \cdot \sum_{i=1}^{n} S_{i}+n \psi_a \\
& =\sum_{i=1}^{n}\left\{\frac{S_{i}}{q}(T(X, Y)-\psi_a)\right\}.
\end{align*}
Therefore,
\begin{align*}
\sqrt{n}\{\hat{\psi_a}(h_{\text{fix}}(X))-\psi_a\} & \rightarrow \frac{1}{\sqrt{n}} \sum_{i=1}^{n}\left[\frac{S_{i}}{q}((X, Y)T-\psi_a)\right], 
\end{align*}
and
\begin{align*}
\AV(\hat{\psi_a}(h_{\text{fix}}(X))) & =\operatorname{Var}\left[\frac{S}{q}(T(X, Y)-\psi_a)\right] \\
& =\E\left[\frac{S}{q^{2}}(T(X, Y)-\psi_a)^{2}\right]-E^{2}\left[\frac{S}{q}(T(X, Y)-\psi_a)\right] \\
& =\E\left[\frac{S}{q^{2}}(T(X, Y)-\psi_a)^{2}\right] \\
& =\E\left[\frac{S}{q^{2}} T(X, Y)^{2}\right]+\E\left[\frac{S}{q^{2}} \psi_a^{2}\right]-2 \E\left[\frac{S}{q^{2}} T(X, Y) \psi_a\right]  \\
& =\frac{1}{q} \E\left[T(X, Y)^{2} \mid S=1\right]+\frac{\psi_a^{2}}{q}-2 \frac{\psi_a}{q} \cdot \psi_a \\
& =\frac{1}{q} \E\left[T(X, Y)^{2} \mid S=1\right]-\frac{\psi_a^{2}}{q}.
\end{align*}
Now we take a closer look at $\E\left[T(X, Y)^2 \mid S=1\right]$.
\begin{align*}
&\E\left[T(X, Y)^2 \mid S=1\right]\\
=&\E\left[\left\{\frac{A}{e_a(X)}(Y-h_{\text{fix}}(X))+h_{\text{fix}}(X)\right\}^{2} \Bigg\rvert\ S=1\right]\\
=&\E\Bigg[\frac{A}{e_a(X)^{2}}\left(Y^{2}+h_{\text{fix}}(X)^{2}-2 Y h_{\text{fix}}(X)\right)+h_{\text{fix}}(X)^{2}+2 \frac{A}{e_a(X)}(Y-h_{\text{fix}}(X)) h_{\text{fix}}(X) \Bigg\rvert\ S=1\Bigg]\\
=&\E\Bigg[\big. A \cdot \frac{1-e_a(X)}{e_a(X)^{2}}\left(\frac{Y^{2}}{1-e_a(X)}+\frac{h_{\text{fix}}(X)^{2}}{1-e_a(X)}-2 \frac{Y h_{\text{fix}}(X)}{1-e_a(X)}\right)+A \cdot \frac{1-e_a(X)}{e_a(X)^{2}} \cdot \frac{e_a(X)^{2}}{A(1-e)} h_{\text{fix}}(X)^{2}+\\
&2 A \cdot \frac{1-e_a(X)}{e_a(X)^{2}} \frac{e_a(X)^{2}}{A(1-e_a(X))} \frac{A}{e_a(X)}(Y-h_{\text{fix}}(X)) h_{\text{fix}}(X) \Bigg\rvert\ S=1\Bigg]\\
=&\E\Bigg[A \cdot \frac{1-e_a(X)}{e_a(X)^{2}}\Bigg\{\frac{Y^{2}}{1-e_a(X)}+\frac{h_{\text{fix}}(X)^{2}}{1-e_a(X)}-\frac{2}{1-e_a(X)} Y h_{\text{fix}}(X)+\frac{e_a(X)^{2}}{A(1-e_a(X))} h_{\text{fix}}(X)^{2}+\\
&\frac{2 e_a(X)}{1-e_a(X)} Y h_{\text{fix}}(X)-\frac{2 e_a(X)}{1-e_a(X)} h_{\text{fix}}(X)\Bigg\} \Bigg\rvert\ S=1\Bigg]\\
=&\E\left[\left.A \cdot \frac{1-e_a(X)}{e_a(X)^{2}}\left\{\frac{1}{1-e_a(X)} Y^{2}+\frac{(A-e_a(X))^{2}}{A(1-e_a(X))} h_{\text{fix}}(X)^{2}-Y h_{\text{fix}}(X)\right\} \right\rvert\, S=1\right]\\
=&\E\left[A \cdot \frac{1-e_a(X)}{e_a(X)^{2}}\left\{\left.(Y-h_{\text{fix}}(X))^{2}+\frac{e_a(X)}{1-e_a(X)} Y^{2}+\frac{e_a(X)^{2}-A e_a(X)}{A(1- e_a(X))} h_{\text{fix}}(X)^{2} \right\rvert\, S=1\right]\right.\\
=&\E\left[\left.A \cdot \frac{1-e_a(X)}{e_a(X)^{2}}(Y-h_{\text{fix}}(X))^{2} \right\rvert\, S=1\right]+\E\left[\left.\frac{A}{e_a(X)} Y^{2} \right\rvert\, S=1\right]+\\
&\E\left[\left.\left(1-\frac{A}{e_a(X)}\right) h_{\text{fix}}(X)^{2} \right\rvert\, S=1\right]\\
=&\E\left[\left.A \cdot \frac{1-e_a(X)}{e_a(X)^{2}}(Y-h_{\text{fix}}(X))^{2} \right\rvert\, S=1\right]+\E\left[\left.\E\left[\left.\frac{A}{e_a(X)} Y^{2} \right\rvert\, X, Y^{a} \cdot S=1\right] \right\rvert\, S=1\right]+\\
&\E\left[\left.\E\left[\left.\left(1-\frac{A}{e_a(X)}\right) h_{\text{fix}}(X)^{2} \right\rvert\, X, S=1\right] \right\rvert\, S=1\right]\\
=&\E\left[\Pr[A=a \mid S=1] \frac{1-e_a(X)}{e_a(X)^{2}}(Y-h_{\text{fix}}(X))^{2} \Bigg\rvert\, S=1 \right]
+\\
&\E\left[\left.\E\left[\left.\frac{A}{e_a(X)}\left(Y^{a}\right)^{2} \right\rvert\, X, Y^{a} , S=1\right] \right\rvert\, S=1\right]\\
=&\E\left[\Pr[A=a \mid S=1] \frac{1-e_a(X)}{e_a(X)^{2}}(Y-h_{\text{fix}}(X))^{2} \Bigg\rvert\, S=1 \right]+\E\left[\left(Y^{a}\right)^{2} \mid S=1\right].
\end{align*}
The last term equals
\[
\E\left[\left(Y^{a}\right)^{2} \mid S=1\right]=\operatorname{Var}\left[Y^{a} \mid S=1\right]+\E^2\left[Y^{a} \mid S=1\right]=\operatorname{Var}\left[Y^{a} \mid S=1\right]+\psi_a^{2}.
\]
Therefore, $
\AV(\hat{\psi_a}(h_{\text{fix}}(X))) =\frac{1}{q}\{\mathcal{C}_a+f_a(h_{\text{fix}}(X), Y)\}$, where $\mathcal{C}_a=\operatorname{Var}\left[Y^{a} \mid S=1\right]$, and 
$$f_a(h_{\text{fix}}(X), Y)=\E\left[\Pr[A=a \mid S=1] \frac{1-e_a(X)}{e_a(X)^{2}}(Y-h_{\text{fix}}(X))^{2} \Bigg\rvert\, S=1 \right].$$ 
This completes the proof.

We also note $\AV(\hat{\psi_a}(h_{\text{fix}}(X)))$ can be decomposed in other ways. For example,
\begin{align*}
& \E\left[T(X, Y)^{2} \mid S=1\right] \\
= & \E\left[\left.\left\{\frac{A}{e_a(X)}(Y-h_{\text{fix}}(X))+h_{\text{fix}}(X)\right\}^{2} \right\rvert\, S=1\right] \\
= & \E\left[\left.\frac{A^{2}}{e_a(X)^{2}}\left(Y^{2}+h_{\text{fix}}(X)^{2}-2 Y h_{\text{fix}}(X)\right)+h_{\text{fix}}(X)^{2}+2 \frac{A}{e_a(X)}(Y-h_{\text{fix}}(X)) h_{\text{fix}}(X) \right\rvert\, S=1\right] \\
= & \E\Bigg[\frac{1}{e_a(X)^{2}}\{A Y^{2}+A h_{\text{fix}}(X)^{2}-2 A Y h_{\text{fix}}(X)+e_a(X)^{2} h_{\text{fix}}(X)^{2}+\\
&2 A e_a(X)(Y-h_{\text{fix}}(X)) h_{\text{fix}}(X) \}\Bigg\rvert\,  S=1\Bigg]. \\
= & \E\left[\frac{1}{e_a(X)^{2}}\left\{A^{2} Y^{2}+(A-e_a(X))^{2} h_{\text{fix}}(X)^{2}-2 A(1-e_a(X)) Y h_{\text{fix}}(X)\right\} \Bigg\rvert\, S=1\right]. \\
= & \E\left[\left.\frac{1}{e_a(X)^{2}}\{A Y-(A-e_a(X)) h_{\text{fix}}(X)\}^{2} \right\rvert\, S=1\right] \\
\end{align*}
Therefore, $
\AV(\hat{\psi_a}(h_{\text{fix}}(X))) =\frac{1}{q}\{\mathcal{C}'_a+f_a'(h_{\text{fix}}(X), Y)\}$, where $\mathcal{C}'_a=-\psi_a^2$, and 
$$f'_a(h_{\text{fix}}(X), Y)=\E\left[\left.\frac{1}{e_a(X)^{2}}\{A Y-(A-e_a(X)) h_{\text{fix}}(X)\}^{2} \right\rvert\, S=1\right].$$ 
We opted for the $f_a(h_{\text{fix}}(X), Y)$ because it is easier to be optimized.
\end{proof}
\subsection{Regularity conditions for M-estimation}\label{app:regularity}
Denote the stack of estimating equations, and their derivative by $\boldsymbol{m}(O_i; \boldsymbol{\boldsymbol{\widetilde \theta}})$ and $\boldsymbol{m'}(O_i; \boldsymbol{\boldsymbol{\widetilde \theta}})=\partial/\partial\boldsymbol{\boldsymbol{\widetilde \theta}} \boldsymbol{m}(O_i; \boldsymbol{\boldsymbol{\widetilde \theta}})$, respectively. Denote the set of possible parameter values of $\boldsymbol{\theta}$ by $\boldsymbol{\Theta}$. We list the regularity conditions for the M-estimators to be consistent and asymptotically normal~\citep{newey1994large}.
\begin{suppthmcond}
    Suppose $\boldsymbol{\theta}=\{\boldsymbol{\theta}_1, \boldsymbol{\theta}_2\}$ and let $\boldsymbol{m}_1(O_i; \boldsymbol{\boldsymbol{\theta}_1})$ be the estimating equation for $\boldsymbol{\theta}_1$. $1/n \sum_{i=1}^{n}\boldsymbol{m}_1(O_i; \boldsymbol{\boldsymbol{\theta}_1})$ and $1/n \sum_{i=1}^{n}\boldsymbol{m}(O_i; \boldsymbol{\boldsymbol{\widehat \theta}_1, \boldsymbol{\theta}_2})$ converge to 0 in probability for any partition of $\boldsymbol{\theta}$.
\end{suppthmcond}

\begin{suppthmcond}
    $\boldsymbol{\theta}$ is in the interior of $\boldsymbol{\Theta}$.
\end{suppthmcond}
\begin{suppthmcond}
    $\E[\boldsymbol{m}(O_i; \boldsymbol{\boldsymbol{\theta}})]$ is continuous and $\text{sup}_{\boldsymbol{\theta}\in\boldsymbol{\Theta}}||1/n\sum_{i=1}^{n}\boldsymbol{m}(O_i; \boldsymbol{\boldsymbol{\theta}})-\E[\boldsymbol{m}(O_i; \boldsymbol{\boldsymbol{\theta}})]||$ converges to 0 in probability.
\end{suppthmcond}
\begin{suppthmcond}
    $\E[\boldsymbol{m}(O_i; \boldsymbol{\boldsymbol{\theta}})]=0$ and $\E[\boldsymbol{m}^2(O_i; \boldsymbol{\boldsymbol{\theta}})]<\infty$.
\end{suppthmcond}
\begin{suppthmcond}
    $\E[\text{sup}_{\boldsymbol{\theta}}||\boldsymbol{m'}(O_i; \boldsymbol{\boldsymbol{\theta}})||]<\infty$.
\end{suppthmcond}
% \begin{suppthmcond}
%     $\boldsymbol{m}(O_i; \boldsymbol{\boldsymbol{\theta}})$ converges almost surely to zero as $n$ goes to infinity.
% \end{suppthmcond}
\begin{suppthmcond}
There is a neighborhood of $\boldsymbol{\theta}$ on which with probability one $-\boldsymbol{m}(O_i; \boldsymbol{\boldsymbol{\widetilde \theta}})$ is continuously differentiable; and $-\boldsymbol{m'}(O_i; \boldsymbol{\boldsymbol{\widetilde \theta}})$ converges uniformly to a non-stochastic limit which is non-singular at $\boldsymbol{\widetilde \theta}$.
\end{suppthmcond}
% Under these two conditions, $\boldsymbol{\boldsymbol{\widehat \theta}}$ is a strongly consistent estiamtor of $\boldsymbol{\boldsymbol{\theta}}$ (Theorem 2 in \citep{yuan1998asymptotics}).
\begin{suppthmcond}
$B(\boldsymbol{\theta})=\E[\boldsymbol{m}(O_i; \boldsymbol{\boldsymbol{\theta}})\boldsymbol{m}(O_i; \boldsymbol{\boldsymbol{\theta}}^{\top})]$ is non-singular and $\sqrt{n}\boldsymbol{m}(O_i; \boldsymbol{\boldsymbol{\theta}})$ converges in distribution to $\mathcal{N}(0, B(\boldsymbol{\theta}))$.
\end{suppthmcond}
% Under these three conditions, $\boldsymbol{\boldsymbol{\widehat \theta}}$ is asymptotically normal (Theorem 4 in \citep{yuan1998asymptotics}).
\subsection{Proof of Lemma~\ref{lemma:reweite}}\label{app:lemma:reweite}

Before we present the proof of Lemma~\ref{lemma:reweite}, we need to prove the following auxilliary lemma. 
\begin{suppthmlem}\label{lemma:conditional_independence}
    Under conditions 1-5 we have that $Y \indep S \mid ( X, A=0 ) $.
\end{suppthmlem}
\begin{proof}
    First, due to the condition of no treatment variation in $\{S=0\}$, we have $Y^a\indep A\mid (X, S=0)$ because $A$ is a constant when $S=0$ (alternatively, in case there is treatment variation in $\{S=0\}$, we can replace condition~5 by directly invoking $Y^a\indep A \mid (X,S=0)$ from condition~5'). Thus, combining this with that $Y^a\indep A\mid (X, S=1)$ (condition 2), we have $Y^a\indep A\mid (X, S)$. Combining this with condition 4, $Y^a\indep S \mid X$, we have $Y^a\indep (A,S)\mid X$. This condition implies $Y^a \indep S\mid (X, A)$, which follows from the weak union property of conditional independence.
    Finally, using condition 1 (consistency), we have $Y \indep S\mid (X, A=0)$.
\end{proof}

Here follows the proof of Lemma~\ref{lemma:reweite}.
\begin{proof}
We will express \eqref{eq:AV_fix} as a quantity that is not conditional on $\{S=1\}$.
We denote $l(X,Y)=\dfrac{\Pr[A=0|S=1]e_1(X)}{e_0^2(X)} \{Y-h(X)\}^2$ and have
\begin{align*}
         f_a(h(X), Y)&=\E\left[\;l(X,Y)\mid S=1, A=0 \right]\\
        & = \E\left[\frac{\mathbf{1}(S=1)}{\Pr[S=1\mid A=0]}\;l(X,Y)\mid A=0\right] \\
        & = \E\left[ \E\left[\frac{\mathbf{1}(S=1)}{\Pr[S=1\mid A=0]}\;l(X,Y) \mid X, A=0\right]\mid A=0\right] \\
        &= \E\left[ \E\left[\frac{\mathbf{1}(S=1)}{\Pr[S=1\mid A=0]}\mid X, A=0\right]\E\left[\;l(X,Y) \mid X, A=0\right]\mid A=0 \right] \\
        &= \E\left[ \frac{\Pr[S=1\mid X, A=0]}{\Pr[S=1\mid A=0]}\E\left[\;l(X,Y) \mid X, A=0\right]\mid A=0 \right] \\
        & =\E\left[\E\left[\frac{\Pr[S=1\mid X, A=0]}{\Pr[S=1\mid A=0]}\;l(X,Y) \mid X, A=0\right]\mid A=0 \right] \\
        & = \E\left[\frac{\Pr[S=1\mid X, A=0]}{\Pr[S=1\mid A=0]}\;l(X,Y)\mid A=0 \right] 
\end{align*}
where the fourth equation follows from that $Y\indep S \mid (X, A=0)$ according to Lemma \ref{lemma:conditional_independence} and the fifth equation follows from that  $\E[\mathbf{1}(S=1)\mid X,A=0]=\Pr[S=1\mid X, A=0]$.
\end{proof}

\subsection{Proof of Theorem \ref{them:efficient}}\label{app:them:efficient}
\begin{proof}
     It is straightforward to verify that the estimating equations in~\eqref{eq:stack_psi} satisfy regularity conditions (listed in Appendix \ref{app:regularity}), and thus $\boldsymbol{\widehat \theta}$ is asymptotically multivariate normal \citep{boos2013essential}. The consistency and asymptotic normality of $\widehat \psi_{0}(\widehat e_0, \widehat h^{*})$ are direct results from the M-estimation theories \citep{stefanski2002calculus} and Lemmas \ref{lemma:fix} and \ref{lemma:reweite}.
\end{proof}
% \begin{proof}
% % Because $\widehat \phi_0$ is a doubly robust estimator \citep{robins1995analysis}, $\widehat \psi_{0}(\widehat e_0, h^{*})$ is consistent under conditions 1-3 when $\widehat e_0$ is correctly specified, regardless whether $h^{*}$ is the correctly specified function for $\E[Y|X, S=1, A=0]$.

% Because the $\boldsymbol{\widehat \theta}$ is estimated through M-estimation, when the regularity conditions hold, $\boldsymbol{\widehat \theta}$ is consistent and asymptotically multivariate normal \citep{stefanski2002calculus}.

% By Lemmas \ref{lemma:fix} and \ref{lemma:reweite}, the $h^{*}(X)$, which is the solution to $\sum_{i=1}^{n}m_{\psi}(\widetilde \psi)$, is the $h(X)$ that minimizes the asymptotic variance of $\widehat \psi_{0}(\widehat e_0, h)$, so that $\widehat \psi_{0}(\widehat e_0, h^{*})$ minimizes the asymptotic variance of $\widehat \psi_{0}(\widehat e_0, h)$ for a fixed $h(X)$. 
%\end{proof}

\subsection{Obtain $\{\widehat \psi_0(\widehat h^{*}), \widehat \psi_1( \widehat g_1)\}$ via M-estimation}\label{app:MforTau}
We already described how to obtain $\widehat \psi_0( \widehat h^{*})$ by M-estimation. Here, we will describe how to obtain $\{\widehat \psi_0( \widehat h^{*}), \widehat \psi_1( \widehat g_1)\}$ by M-estimation. Compared to the estimation for $\psi_0$, estimating $\{\psi_0, \psi_1\}$ jointly requires two additional estimating equations, for estimating the parameters in $g_1(X)$, and for obtaining $\widehat \psi_0( \widehat h^{*})$, respectively.

Denote $g_1(X)$ by $g_1(X; \zeta)$, and let $\boldsymbol{\theta'}=\{ q,  \beta,  \gamma,  \psi_{0}, \zeta, \psi_1\}$ be a vector of smooth finite dimensional target parameters.
We propose to estimate $\boldsymbol{\theta'}$ by finding the $\boldsymbol{\widetilde \theta}'$ that solves the following joint estimating equation
\begin{equation*}
\begin{pmatrix}
    m_q(O_i; \widetilde q)\\
    m_{\eta_0}(O_i; \widetilde \beta)\\
    m_{h^{*}}(O_i; \widetilde \beta, \widetilde \gamma)\\
    m_{\psi_0}(O_i; \widetilde q,, \widetilde \gamma, \widetilde \psi_0)\\
    m_{g_1}(O_i; \widetilde \zeta)\\
    m_{\psi_1}(O_i; \widetilde q,, \widetilde \zeta, \widetilde \psi_1)\\
    % m_{\tau(\widehat h^{*})}(O_i; \widetilde \psi_0, \widetilde \psi_1, \widetilde \tau(\widehat h^{*}))\\
\end{pmatrix}
=\mathbf{0},
\end{equation*}
where the first four estimating equations are the same four estimating equations in \eqref{eq:stack_psi}.

The estimating equation for $g_1(X; \widetilde \zeta)$ depends on the outcome. For example, when the outcome $Y$ is binary, $m_{g_1}(O_i; \widetilde \zeta)$ is the score of the logistic regression models for $g_1(X; \zeta)$. To obtain $\widehat \psi_{1}( \widehat g_1)$, according to the construction of the estimator, we define 
$$
m_{\psi_1}(O_i; \widetilde q, \widetilde \zeta, \widetilde \psi_1)=\frac{1}{\widetilde q}S_i\left[\frac{\mathbf{1}(A_i=1)}{e_{1}(X_i)}\{Y_i- g_1(X_i, \widetilde \zeta)\} +   g_1(X_i, \widetilde \zeta) -\widetilde \psi_1\right].
$$
% To obtain $\widehat \tau(\widehat h^{*})$, we define 
% $$
% m_{\tau(h^{*})}(O_i; \widetilde \psi_0, \widetilde \psi_1, \widetilde \tau)=\widetilde \psi_1-\widetilde \psi_0-\widetilde \tau(\widehat h^{*}).
% $$
It is straightforward to verify that the above estimating equations satisfy regularity conditions (listed in Appendix \ref{app:regularity}).

\subsection{Proof of Lemma~\ref{lem:combined_consistency}}\label{app:lem:combined_consistency}
\begin{proof}
    If $\widehat \tau(\widehat g)$ and $\widehat \tau(\widehat h^{*})$ are obtained via joint M-estimation (the details are given in Appendix \ref{app:MforTauLambda}) then they are are asymptotically bivariate normal \citep{boos2013essential}. Because a linear combination of two bivariate normal distributed random variables is normally distributed, $\widehat \tau(\widehat \lambda^{*})$ is asymptotically normal. As both $\widehat \tau(\widehat g)$ and $\widehat \tau(\widehat h^{*})$ are consistent estimators, it also follows that their linear combination is consistent.
\end{proof}

\subsection{Obtain $\{\widehat \tau(\widehat g), \widehat \tau(\widehat h^{*})\}$ via M-estimation}\label{app:MforTauLambda}
We already described the how to obtain $\{\widehat \psi_0(\widehat h^{*}), \widehat \psi_1( \widehat g_1)\}$ via M-estimation in Appendix \ref{app:MforTau}. Here, we describe how to obtain $\{\widehat \tau(\widehat g), \widehat \tau(\widehat h^{*})\}$ via M-estimation for a given $\widehat \lambda^{*}$. Observe that 
\begin{align*}
    \widehat \tau(\widehat \lambda^{*})=&
    \widehat \lambda^{*}\widehat \tau(\widehat g)+(1-\widehat \lambda^{*})\widehat \tau(\widehat h^{*})\\
    =&\widehat \lambda^{*}\{\widehat \psi_1( \widehat g_1)-\widehat \psi_0(\widehat g_0)\}+(1-\widehat \lambda^{*})\{\widehat \psi_1( \widehat g_1)-\widehat \psi_0( \widehat h^{*})\}.
\end{align*}
Compared to the M-estimation described in Appendix \ref{app:MforTau}, obtaining $\{\widehat \tau(\widehat g), \widehat \tau(\widehat h^{*})\}$ via M-estimation requires four additional estimating equations, for estimating the parameters in $\widehat g_0(X)$, and for obtaining $\widehat \tau(\widehat h^{*})$, $\widehat \psi_0( \widehat g_0)$, and  $\widehat \tau(\widehat g)$, respectively.

Denote $g_0(X)$ by $g_0(X; \iota)$, and let 
$
\boldsymbol{ \theta''}=\{ q,  \beta,  \gamma,  \psi_{0},  \zeta,  \psi_1, \tau(\widehat h^{*}), \iota, \tau(\widehat g)\}
$
be a vector of smooth finite dimensional targeted parameters parameters.
We propose to estimate $\boldsymbol{\theta''}$ by finding the $\boldsymbol{\widetilde \theta''}$ that solves the following joint estimating equation
\begin{equation*}
\begin{pmatrix}
    m_q(O_i; \widetilde q)\\    
    m_{\eta_0}(O_i; \widetilde \beta)\\
    m_{h^{*}}(O_i; \widetilde \beta, \widetilde \gamma)\\
    m_{\psi_0}(O_i; \widetilde q,  \widetilde \gamma, \widetilde \psi_0)\\
    m_{g_1}(O_i; \widetilde \zeta)\\
    m_{\psi_1}(O_i; \widetilde q,  \widetilde \zeta, \widetilde \psi_1)\\
    m_{\tau(h^{*})}(O_i; \widetilde \psi_1,  \widetilde \psi_0, \widetilde \tau(\widehat h^{*}))\\
    m_{g_0}(O_i; \widetilde \iota)\\
    m_{\psi_0^{'}}(O_i; \widetilde q,  \widetilde \iota, \widetilde \psi_0^{'})\\
    m_{\tau(\widehat g)}(O_i; \widetilde \psi_1, \widetilde \psi_0^{'}, \widetilde \tau(\widehat g))\\
    % m_{\tau(\lambda^{*})}(O_i; \widetilde \tau(g), \widetilde \tau(h^{*}), \widetilde \tau(\lambda^{*}))\\
\end{pmatrix}
=\mathbf{0},
\end{equation*}
where the first six estimating equations are the same six estimating equations in Appendix \ref{app:MforTau}.

The estimating equation for $g_0(X; \iota)$ depends on the outcome. For example, when the outcome $Y$ is binary $m_{g_0}(O_i; \widetilde \iota)$ is the score of the logistic regression models for $g_0(X; \widehat \iota)$. To obtain $\widehat \psi_0'$ (which is the estimate of $\psi_0$ using the trial-only estimator),  $\widehat \tau(\widehat g)$, and $\widehat \tau(\widehat \lambda^{*})$, we define 
\begin{align*}
& m_{\tau(h^{*})}(O_i; \widetilde \psi_1,  \widetilde \psi_0, \widetilde \tau(\widehat h^{*}))=\widetilde \psi_1- \widetilde \psi_0-\widetilde \tau(\widehat h^{*}).\\
&m_{\psi_0^{'}}(O_i; \widetilde q, \widetilde \iota, \widetilde \psi_0^{'})=\frac{1}{\widetilde q}S_i\left[\frac{\mathbf{1}(A_i=0)}{1-e_{1}(X_i)}\{Y_i- \widetilde g_0(X_i, \widetilde \iota)\} +  \widetilde g_0(X_i, \widetilde \iota) -\widetilde \psi_0^{'}\right]\\
&m_{\tau(\widehat g)}(O_i; \widetilde \psi_1, \widetilde \psi_0^{'}, \widetilde \tau(\widehat g))=\widetilde \psi_1- \widetilde \psi_0^{'}-\widetilde \tau(\widehat g).\\
% &m_{\tau(\lambda^{*})}(O_i; \widetilde \tau(g), \widetilde \tau(h^{*}), \widetilde \tau(\lambda^{*}))=\widehat \lambda^{*}\widetilde \tau(g)+(1-\widehat \lambda^{*})\widetilde \tau(h^{*})-\tau(\widehat \lambda^{*}).
\end{align*}
% Because $\widetilde \psi_{\cdot}(\cdot, \cdot)$ can be written as a function of data and parameters according to the estimating equations $m_{\psi_{\cdot}(\cdot, \cdot)}=0$, the estimating equations for $\tau_{\widehat h^{*}}$ and $\tau_{\widehat g}$ are valid estimating equations (that satisfy the regularity conditions listed in Appendix \ref{app:regularity}). 
It is straightforward to verify that all the above estimating functions satisfy regularity conditions (listed in Appendix \ref{app:regularity}).

\subsection{Conditions for $\lambda^*$ to be well-defined}\label{app:existlambda}
When $\sigma^2_g + \sigma^2_{h^*} -  \sigma_{g,h^*} =0$, $\lambda^*$ is not well defined. We will prove here that this pathological case only happens when $\widehat g$ and $\widehat h^*$ converge to the same limit.

Given that the vector $\{\widehat \tau(\widehat g), \widehat \tau(\widehat h^{*})\}$ is jointly asymptotic normal, we can write
$$
\sigma^2_g + \sigma^2_{h^*} -  \sigma_{g,h^*} =0 \iff \AV[\widehat \tau(\widehat g)  - \widehat \tau (\widehat h^*)] = \AV[\widehat \tau(g^\dagger)  - \widehat \tau (h^\dagger)] =  0,
$$
where $g^\dagger=\{g^\dagger_0, g^\dagger_1\}$ and $h^\dagger$ are the limits of the (possibly misspecified) outcome functions, that is for $a \in \{0,1\}$ it holds that $||\widehat g_a - g_a^\dagger||_2 = o_P(1)$ and $||\widehat h^* - h^\dagger||_2 = o_P(1)$. 

Now, we rewrite the difference of these two estimators as follows: 
\begin{align*}
\widehat \tau(g^\dagger)  - \widehat \tau(h^\dagger)  = \frac{1}{n}\sum_{i=1}^n \Gamma_i, ~~\text{where}~~\Gamma_i = \frac{S_i}{P(S=1)}  \left[ \left(\frac{1-A_i}{e_0(X_i)}-1 \right)\{h^\dagger(X_i)-g_0^\dagger(X_i)\}\right].
\end{align*}
With the shorthand above we can rewrite the asymptotic variance of the difference as
\begin{align*}
\AV[\widehat \tau(g^\dagger)  - \widehat \tau (h^\dagger)] = \operatorname{Var}  [\Gamma_i] = \operatorname{Var} \left [ \frac{S}{P(S=1)}   \left(\frac{1-A}{e_0(X)}-1 \right)\{h^\dagger(X)-g_0^\dagger(X)\}\right].
\end{align*}
Finally,  we can conclude that
\begin{align*}
\sigma^2_g + \sigma^2_{h^*} -  \sigma_{g,h^*} =0 &\iff 
\AV[\widehat \tau(g^\dagger)  - \widehat \tau (h^\dagger)] = 0\\&\iff \operatorname{Var} \left [ \frac{S}{P(S=1)}   \left(\frac{1-A}{e_0(X)}-1 \right)\{h^\dagger(X)-g_0^\dagger(X)\}\right] =0\\
&\iff \frac{S}{P(S=1)}   \left(\frac{1-A}{e_0(X)}-1 \right)\{h^\dagger(X)-g_0^\dagger(X)\} =0,~\mathrm{ almost~surely}\\
&\iff h^\dagger(X)=g_0^\dagger(X),~\mathrm{ almost~surely}.
\end{align*}
Hence, $\lambda^*$ is ill-defined if and only if $\widehat h^*$ and $\widehat g$ converge to the same  limit (up to differences on some measure zero sets).

\subsection{Proof of Theorem \ref{thm_combine_optimal}}\label{app:thm_combine_optimal}

%We first introduce two lemmas needed for the main proof.
In what follows we denote convergence in distribution and probability by $\overset{d}{\to}$ and $\overset{p}{\to}$, respectively.

\begin{proof}
If $\widehat \tau(\widehat g)$ and $\widehat \tau(\widehat h^{*})$ are obtained via joint M-estimation (the details are given in Appendix \ref{app:MforTauLambda}) then they are are asymptotically bivariate normal. More formally, we can write 
$$
\sqrt{n} \begin{pmatrix} \widehat \tau(\widehat{h}^*) - \tau \\ \widehat \tau(\widehat{g}) - \tau \end{pmatrix} \xrightarrow{d}Z,~~\text{with}~~ Z \sim  \mathcal{N} \left( \mathbf 0, \Sigma \right)~~\text{and}~~  \Sigma = \begin{pmatrix} \sigma^2_{h^*} & \sigma_{g,h^*} \\\sigma_{g,h^*} & \sigma^2_{g} \end{pmatrix}.
$$
Further, for any $\lambda \in \mathbb R$, the combined estimators is equal to
$$
\widehat \tau(\lambda) = \lambda \widehat\tau(\widehat h^*) + (1-\lambda) \widehat \tau(\widehat g) = (\lambda, 1-\lambda)^\top \begin{pmatrix} \widehat \tau(\widehat{h}^*)  \\ \widehat \tau(\widehat{g})  \end{pmatrix} .
$$
Since by assumption $\widehat \Sigma \xrightarrow{p} \Sigma$, it follows from the continuous mapping theorem that $\widehat \lambda^{*}\overset{p}{\to} \lambda^{*}$. As a consequence, we have from Slutsky's theorem that 
$$
\sqrt n (\widehat \tau(\widehat \lambda^*) - \tau) \xrightarrow{d} (\lambda^*, 1-\lambda^*)^\top Z.
$$
 Therefore, we have asymptotic normality of the combined estimator and moreover, the asymptotic variance is given by:
\begin{align*}
\text{AVar}[\widehat \tau(\widehat \lambda^*)] &= \text{AVar}[\widehat \tau( \lambda^*)] = (\lambda^*, 1-\lambda^*)^\top \Sigma (\lambda^*, 1-\lambda^*) =(\lambda^{*})^2 \sigma^2_{h^*} + (1-\lambda^*)^2 \sigma^2_{g} + 2 \lambda^*(1-\lambda^*) \sigma_{g,h^*} .
\end{align*}
\end{proof}

\section{Discussion on study designs}
\label{app:study_design}

Settings with data from multiple sources can often be categorized as either \textit{nested trial designs} or \textit{non-nested trial designs}~\citep{dahabreh2020extending,dahabreh2021study,li2022generalizing}. For nested trial designs, the target population is well-defined with the trial population nested inside of it; often, the target population corresponds to a census from which trial-eligible individuals are selected. Those in the census who are not selected to participate in the trial correspond to the external population. Meanwhile, in non-nested trial designs the datasets are obtained separately. In this case, we do not know how the datasets were sampled from the target population.

\begin{figure}[t]
    \centering
    \includegraphics[width=0.35\textwidth]{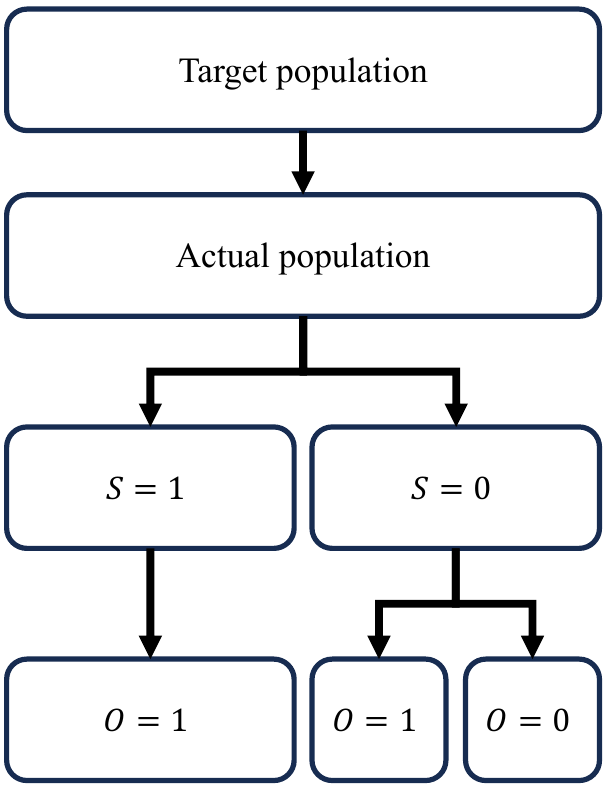}
    \caption{A diagram to conceptualize different study designs.}
    \label{fig:study_design}
\end{figure}

Following the framework in~\cite{dahabreh2021study}, the sampling mechanisms in both the nested and non-nested trial design can be formalized by introducing an indicator variable $O$. The variable $O$ indicates whether an individual from the underlying target population is in the observed data: $\{O=1\}$ for sampled individuals and $\{O=0\}$ for non-sampled individuals. Using the diagram in Figure~\ref{fig:study_design}, we can illustrate how individuals from the underlying target population are first sampled into the actual population before they are divided into two sub-populations; in this case, either the trial population $\{S=1\}$ or the external population $\{S=0\}$. Here, we assume observations are simple random samples from each respective sub-population, meaning that is the sampling probabilities are determined by $\Pr[O=1\mid S=1]$ and $\Pr[O=1\mid S=0]$. Without any loss of generalization, we shall assume that all individuals in the trial sub-population are observed, i.e. $\Pr[O=1\mid S=1]=1$. We let $\Pr[O=1\mid S=0]=u$ for some constant $0<u\leq 1$, for nested trial designs is $u$ known but for non-nested designs is $u$ unknown. 

Due to the simple random sampling, the observation indicator $O$ is independent of the other variables conditioned on the sub-population $S$; that is $O\indep (X, A,Y^1,,Y^0) \mid S$. This means $\E[Y^1-Y^0\mid S=s]=\E[Y^1-Y^0\mid S=s, O=1]$, which implies that the average treatment effect in both sub-populations is still identifiable from the observed individuals. However, we see that $\E[Y^1-Y^0]\neq \E[Y^1-Y^0 \mid O=1]$ can happen. The average treatment population on the target population is thus not identifiable unless we have a nested trial design, in which case we can get around this because $u$ is known~\citep{dahabreh2021study}. Still, $\E[Y^1-Y^0 \mid O=1]$ can be interpreted as the average treatment effect on a mixture of the trial and external population that excludes all unobserved sub-populations~\citep{li2023improving}.

We denote $n$ as the number of observed individuals $(S,X,A,Y,O=1)$ and $N$ as the number of total individuals in the actual population $(OS,OX,OA,OY,O)$. For obvious reasons, $N$ is unknown to us. We assume that ratio $n_s/n \rightarrow q_s>0$, for $S=0,1$,  as $n\rightarrow\infty$. 

\subsection{Identifiability of~\eqref{eq:rewrite} in non-nested study designs}

To minimize~\eqref{eq:rewrite}, we need to estimate the study participation model $\eta_0(X)=P[S=1\mid X, A=0]$. While $\eta_0$ is identifiable from the observed data in a nested design, this is not necessarily the case in a non-nested design. Whereas one could believe this will cause issues for minimizing~\eqref{eq:rewrite}, we shall however show that this does not matter in the end and that~\eqref{eq:rewrite} still is identifiable from observations only; that is, even when everything is conditioned on $\{O=1\}$.

Letting $l(X,Y)=\dfrac{\Pr[A=0|S=1]e_1(X)}{e_0^2(X)} \{Y-h_{fix}(X)\}^2$, the claim of lemma~\ref{lemma:reweite} is that
$$
\E[l(X,Y) \mid A=0, S=1] = \E\left[\frac{\Pr[S=1\mid X, A=0]}{\Pr[S=1\mid A=0]}\frac{\Pr[A=0|S=1]e_1(X)}{e_0^2(X)} \{Y-h_{\text{fix}}(X)\}^2\middle\vert A=0 \right]~.
$$
However, due to simple random sampling of observations, we have
$$
\E[l(X,Y) \mid A=0, S=1]=\E[l(X,Y) \mid A=0, S=1, O=1]~.
$$
Following the same proof as in the lemma, we can then show that
\begin{align*}
        \E&\left[\;l(X,Y)\mid S=1, A=0,O=1\right]
        = \\
        & = \E\left[\frac{\mathbf{1}(S=1)}{\Pr[S=1\mid A=0, O=1]}\;l(X,Y)\mid A=0, O=1\right] \\
        & = \E\left[ \E\left[\frac{\mathbf{1}(S=1)}{\Pr[S=1\mid A=0,O=1]}\;l(X,Y) \mid X, A=0, O=1\right]\mid A=0, O=1\right] \\
        &= \E\left[ \E\left[\frac{\mathbf{1}(S=1)}{\Pr[S=1\mid A=0, O=1]}\mid X, A=0, O=1\right]\E\left[\;l(X,Y) \mid X, A=0, O=1\right]\mid A=0 , O=1\right] \\
        &= \E\left[ \frac{\Pr[S=1\mid X, A=0, O=1]}{\Pr[S=1\mid A=0, O=1]}\E\left[\;l(X,Y) \mid X, A=0, O=1\right]\mid A=0, O=1 \right] \\
        & =\E\left[\E\left[\frac{\Pr[S=1\mid X, A=0, O=1]}{\Pr[S=1\mid A=0, O=1]}\;l(X,Y) \mid X, A=0, O=1\right]\mid A=0, O=1 \right] \\
        & = \E\left[\frac{\Pr[S=1\mid X, A=0. O=1]}{\Pr[S=1\mid A=0, O=1]}\;l(X,Y)\mid A=0, O=1 \right] 
\end{align*}
where the fourth equation follows from that $Y\indep S \mid (X, A=0, O=1)$. This follows from lemma~\ref{lemma:conditional_independence} and that we have simple random sampling. It is easy to show that $O\indep (X. A, Y^1,Y^0) \mid S \Rightarrow O \indep Y \mid (S, X, A=0)$ which combined with $Y\indep S \mid (X, A=0)$ implies that $Y\indep S \mid (X, A=0, O=1)$.  Thus, to conclude, we see in fact that $\E[l(X,Y) \mid A=0, S=1]$ can be expressed as an expectation over quantities of only observed data $\{O=1\}$.

\section{Discussion on using external control data under exchangeability of populations} \label{app:estimation_under_exchangeability}

In settings where data from multiple sources are combined, transportability conditions are often assumed to enable integration. Unlike our approach, existing methods typically construct estimators based on an identification strategy that depends on conditions 4 and 5 (or 5'). While these additional conditions can be controversial, they are often justified alongside conditions 1-3 through testable implications in the observed data. We discuss these issues below and further detail one estimator based on this strategy, which we use in our simulation study and application.

\subsection{Identification assuming exchangeability of populations}

Under conditions 1-5, pooling of trial and external control data can be incorporated in analyses aiming to estimate the potential outcome mean in the population underlying the trial under intervention of the control treatment $A$ to $a=0$, that is, $\E[Y^0\mid S=1]$~\citep{li2023improving,valancius2023causal}. Specifically, we can identify $\E[Y^0\mid S=1]$ with
$$\zeta_0 = \E[ \E[Y\mid X, A=0] \mid S=1].$$
Compared to the identification results using the trial data alone, we no longer condition on $S=0$ in the inner expectation; instead, we pool the trial and external control data. The average treatment effect in the population underlying the trial, $\E[Y^1-Y^0\mid S=1]$, is identified with $\psi_1-\zeta_0$.

\subsection{Testable implications of the additional identifiability conditions}

Conditions~1-5 together have a testable implication in the law of observed data, namely that for each $x$ with positive density in the population
underlying the trial, $f(x,S=1)\neq 0$,
\begin{equation}
    \label{eq:testable_implication}
    H_0 : \E[Y \mid X=x, A=0, S=1] = \E[Y \mid X=x, A=0,S=0].
\end{equation}
To see this, we have from condition~4  that
\begin{equation*}
    \E[Y^0\mid X,S=1]=\E[Y^0\mid X,S=0]~.
\end{equation*}
Then, the above testable implication follows by noting that conditions~1-3 allow us to re-write the left-hand as $\E[Y^0\mid X,S=1]=\E[Y\mid X,A=0,S=1]$.
Meanwhile, the right-hand side can be written as $\E[Y^0\mid X,S=0] =\E[Y^0\mid X,A=0,S=0]=\E[Y\mid X,A=0,S=0]$ where first equality follows from having no treatment variation in $\{S=0\}$ (condition~5) and the second one from consistency (condition~1). 

The testable implication provides a way to evaluate whether conditions~1-5 jointly hold; informally, it may be used to assess compatibility between the trial and external control data. Various methods exist for testing $H_0$, such as parametric likelihood-ratio tests or non-parametric alternatives~\citep{racine2006testing, luedtke2019omnibus}. Complications related to doing a statistical test against $H_0$ and subsequently drawing statistical inferences using the same dataset can be addressed by sample-splitting or accounting for pre-testing when quantifying uncertainty (see, e.g., \cite{rothenhausler2020model, yang2023elastic}).

\subsection{Estimation under exchangeability of populations}
We now present an existing estimator based on the identification strategy presented above, later we refer to it as the pooling estimator in our simulation study and application. \cite{li2023improving} proposed a doublY-robust estimator for $\zeta_0$,
\begin{equation}\label{eq:estimation_assuming_transport}
    \widehat{\zeta}_0 = \left(\sum_{i=1}^n S_i \right)^{-1} \sum_{i=1}^{n} \left[ \left(\frac{S_i (1-A_i) + (1-S_i) \widehat{r}(X_i) }{\widehat{\eta}(X_i) (1-\widehat{e}_0(X_i)) + (1-\widehat{\eta}(X_i))\widehat{r}(X_i)}\widehat{\eta}(X_i)\right)(Y_i - \widehat{g}_0(X_i)) + S_i \widehat{g}_0(X_i)\right]~,
\end{equation}
where $\widehat \eta(X)$ is an estimator for the probability of participation in the trial $\Pr[S=1\mid X]$, $\widehat e_0(X)$ is an estimator for the propensity score, and $\widehat r(X)$ is an estimator for the variance ratio $r(X) \equiv \V[Y^0\mid X,S=1]/\V[Y^0\mid X, S=0]$ comparing the trial population and the population underlying the external data. The estimator of the variance ratio $\widehat r(X)$ controls how much information to ``borrow'' from the external control data; this becomes evident by setting $r(X)=0$ in which case $\widehat{\zeta}_0 = \widehat{\phi}_0$. 

Under conditions~1-5, the estimator $\widehat{\zeta}_0$ is consistent if either the models for estimating ${\eta}(X)$ and ${e}_0(X)$ are correctly specified, or if the model for estimating ${g}_0(X)$ is correctly specified. Furthermore, if all working models are correctly specified, including for ${r}(X)$, $\widehat{\zeta}_0$ is the efficient estimator for the control outcome mean when trial and external control data are available~\citep{li2023improving}. Also, under conditions 4 and 5, one can replace ${g}_0(X)$ with an estimator for $\E[Y\mid X, A=0]$ using both the trial and external control data to borrow more information from the external population. However, if conditions~4 and 5 do not hold (i.e. the external controls are not exchangeable) or the model for estimating ${\eta}(X)$ is misspecified, then $\widehat\zeta_0$ does not have these desirable properties whereas the trial-based estimator $\widehat{\phi}_0$ remains consistent and is the most efficient estimator that ignores the external control data.

\section{Simulation studies}\label{app:simulation_studies}

\subsection{Data-generating process}

For a given $n_1$ and $n_0$, we let $S_i=1$ for $i=1,\dots,n_1$ and $S_i=0$ for $i=n_1+1,\dots, n_1+n_0$. For observations with $S_i=1$, we generated $A_i\sim \text{Bern}(1/2)$; for observations with $S_i=0$, we let $A_i=0$. The covariates $X_i$ were sampled from a $10$-dimensional multivariate Normal distribution $N(\mu_{S_i}, \Sigma)$ where $\mu_s$ depended on the scenario being considered; the diagonal elements of $\Sigma$ were set to one and the off-diagonal elements were set to zero. We generated outcomes according to $Y_i = \sum_{j=1}^{5} \alpha_j X_{i,j} + \sum_{j=1}^{10} \beta_j X_{i,j}^2 + 5 \cdot A_i + \varepsilon_i$ with $\varepsilon_i \sim N(0, 1)$, where we let $\alpha=(\frac{1}{2}, 1, -\frac{1}{2}, 1, -\frac{1}{2})$ and $\beta=(-\frac{1}{4},-1, -\frac{1}{2}, -1, -\frac{1}{2}, \frac{1}{2}, \frac{1}{2}, \frac{1}{2}, \frac{1}{2}, \frac{1}{2})$. 

For scenario A, all parametric working models were correctly specified and we set $\mu_{1}=\mu_{0}=\mathbf{0}$ such that there was no distribution shift for the baseline covariates between the trial population and the population underlying the external control data. Meanwhile, for the more adversial scenario B, working models were misspecified by intentionally omitting the variables with $j=5,\dots,10$ and all second-order terms, and we introduced distribution shift by setting $\mu_{1}=\mathbf{0}$ and $\mu_{0}=\frac{1}{2}\mathbf{1}$.

\subsection{Additional results}

To demonstrate how the relative sizes of the trial and external control groups influence the outcomes, we fixed the number of external controls at $n_0=200$ while varying the size of the trial group. Specifically, two settings were considered: (1) a small trial group with $n_1=50$ and (2) a large trial group with $n_1=200$. 

We compare the following estimators: the unadjusted trial-only, the trial-only augmented inverse probability weighting (AIPW), an estimator that fully pools under exchangeability \citep{li2023improving}, test-then-pool, selective borrowing~\citep{gao2023integrating}, and Bayesian dynamic borrowing~\citep{viele2014use}, alongside the novel estimators proposed.

In scenario A for the small trial, all estimators had negligible bias and primarily differed in terms of variance. Among these, the pooling estimator achieved the lowest variance, followed by test-then-pool, optimized randomization-aware, combined, AIPW, selective borrowing, dynamic borrowing, and finally, the unadjusted trial-only estimator. Meanwhile, in scenario A for the large trial, the variance difference between AIPW (the best trial-only estimator) and estimators that incorporate external controls, excluding dynamic borrowing, became nearly negligible.

In scenario B, a significant increase in bias was observed for pooling, test-then-pool, and dynamic borrowing, regardless of trial size. This bias was slightly more pronounced in the small trial compared to the large trial. For the small trial, the variance ordering remained largely unchanged, except that dynamic borrowing performed comparably to optimized randomization-aware and combined estimators. The increase in AIPW variance in this scenario can be attributed to its sensitivity to misspecification, even though it does not incorporate information from external controls.

\paragraph{Effect of removing weighting in the objective of optimized randomization-aware estimator} In addition to the comparison between estimators, we investigated the importance of the weighting in the objective for $h^*(X)$ in the optimized randomization-aware estimator. This was done by repeating the same experiments as above but finding the minimizer that solves $\argmin_{\tilde{h}} \E\left[\{Y-\widetilde h(X)\}^2\middle\vert A=0 \right]$. As shown in Table~\ref{tab:ablation_study}, the weighting had no impact in the scenario A. However, in the more adversarial scenario B, removing the weighting led to increased bias and variance. This outcome is unsurprising: weighting is necessary for improving inference on a targeted population, particularly when there is a distributional shift between populations and misspecification (see e.g. \citet{shimodaira2000improving}). The slight increase in bias, despite the robustness property of the randomization-aware estimators, could be due to the absence of weighting, which likely exacerbates the misspecification of the outcome model. This phenomenon has been observed to introduce non-negligible bias for doublY-robust estimators with a correctly specified propensity score in small sample sizes~\citep{kang2007demystifying}.

\paragraph{Effect of replacing true propensity score with estimated propensity score} We also explored a scenario where the true propensity score is replaced with an estimated propensity score in the optimized randomization-aware estimator. This investigation was motivated by the fact that, while the probability of treatment is known by design in a trial, estimating it is often preferred to improve efficiency (see, e.g., \citet{williamson2014variance}). However, as shown in Table~\ref{tab:ablation_study}, using an estimated propensity score results in higher variance compared to using the true propensity score. Although this finding might seem to contradict recommendations in the literature, a plausible explanation is that estimating the propensity score introduces additional uncertainty, which destabilizes the minimization of the objective for $h^*(X)$.

\begin{figure}[ht]
    \centering
    \begin{subfigure}[b]{0.45\textwidth}
        \includegraphics[width=\textwidth]{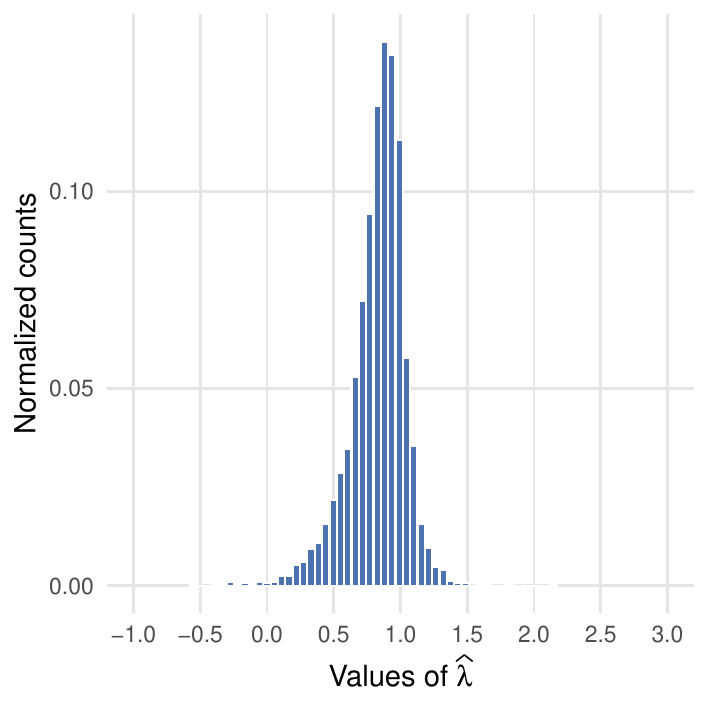}
        \caption{Scenario A}
        \label{fig:lambda_best}
    \end{subfigure}~
    \begin{subfigure}[b]{0.45\textwidth}
        \includegraphics[width=\textwidth]{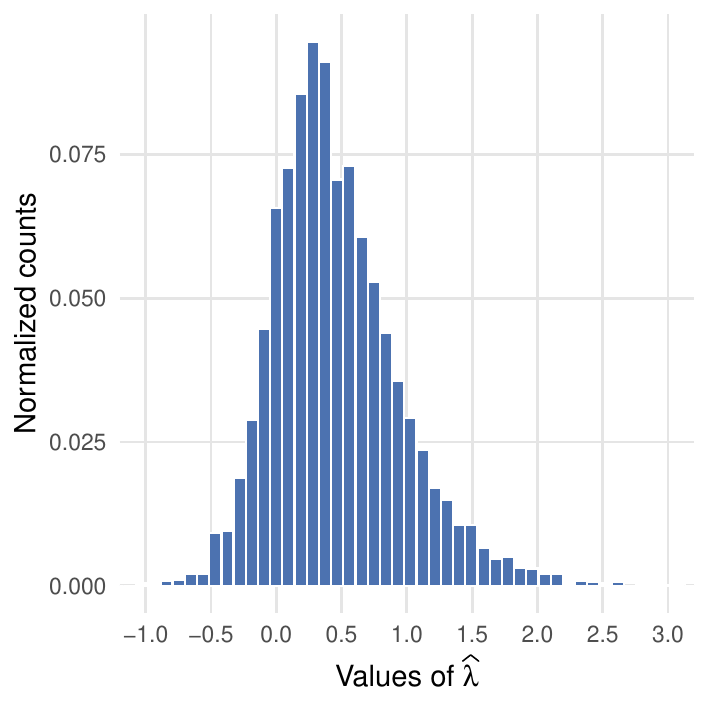}
        \caption{Scenario B}
        \label{fig:lambda_adversarial}
    \end{subfigure}
    \caption{Distribution of $\hat{\lambda}^*$ for the combined estimator in the small trial case. Recall that when $\widehat{\lambda}^*=0$  the combined estimator reduces to AIPW, and when $\widehat{\lambda}^*=1$ it reduces to the optimized randomization-aware estimator. The histogram shows the normalized counts over 5000 repeated experiments.}
    \label{fig:lambda_dist}
\end{figure}

\paragraph{Distribution of $\widehat{\lambda}$ in combined estimator}
To examine the behavior of the combined estimator, we analyzed the distribution of $\widehat{\lambda}$ in the small trial setting. Recall that when $\widehat{\lambda}=0$  the combined estimator reduces to AIPW, and when $\widehat{\lambda}=1$ it reduces to the optimized randomization-aware estimator. In scenario A, where the optimized randomization-aware estimator is expected to perform better, the distribution of $\widehat{\lambda}$ was centered closer to 1  (see Figure~\ref{fig:lambda_best}). Conversely, in scenario B where the optimized randomization-aware is less favored, the distribution shifted closer to 0 (see Figure~\ref{fig:lambda_adversarial}). Interestingly, in both scenarios, $\widehat{\lambda}$ and in some instances, even outside this range. This suggests that the combined estimator frequently would leverage information from both estimators.

\begin{figure}[ht]
    \centering
    \begin{subfigure}[b]{0.4\textwidth}
        \includegraphics[width=\textwidth]{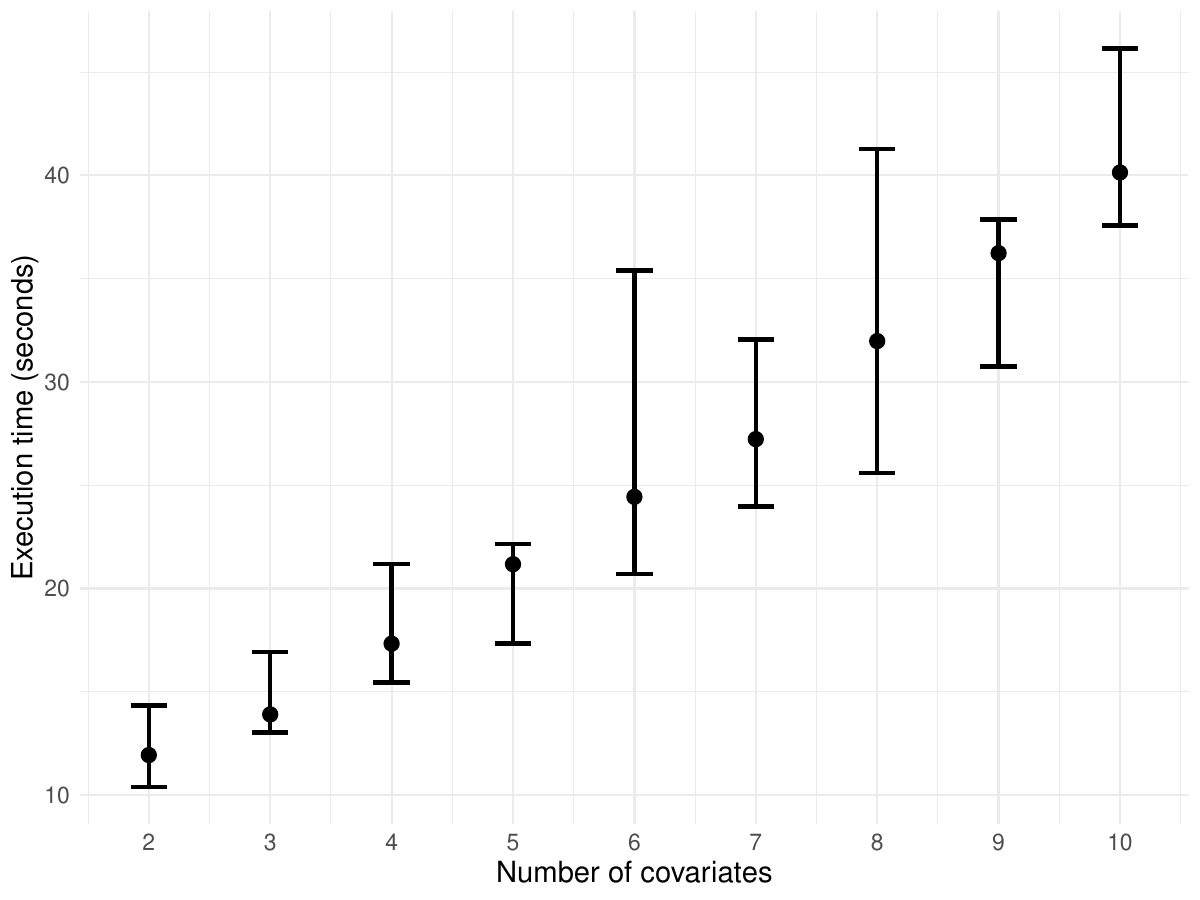}
        \caption{Varying number of parameters}
        \label{fig:vary_sample_size}
    \end{subfigure}~
    \begin{subfigure}[b]{0.4\textwidth}
        \includegraphics[width=\textwidth]{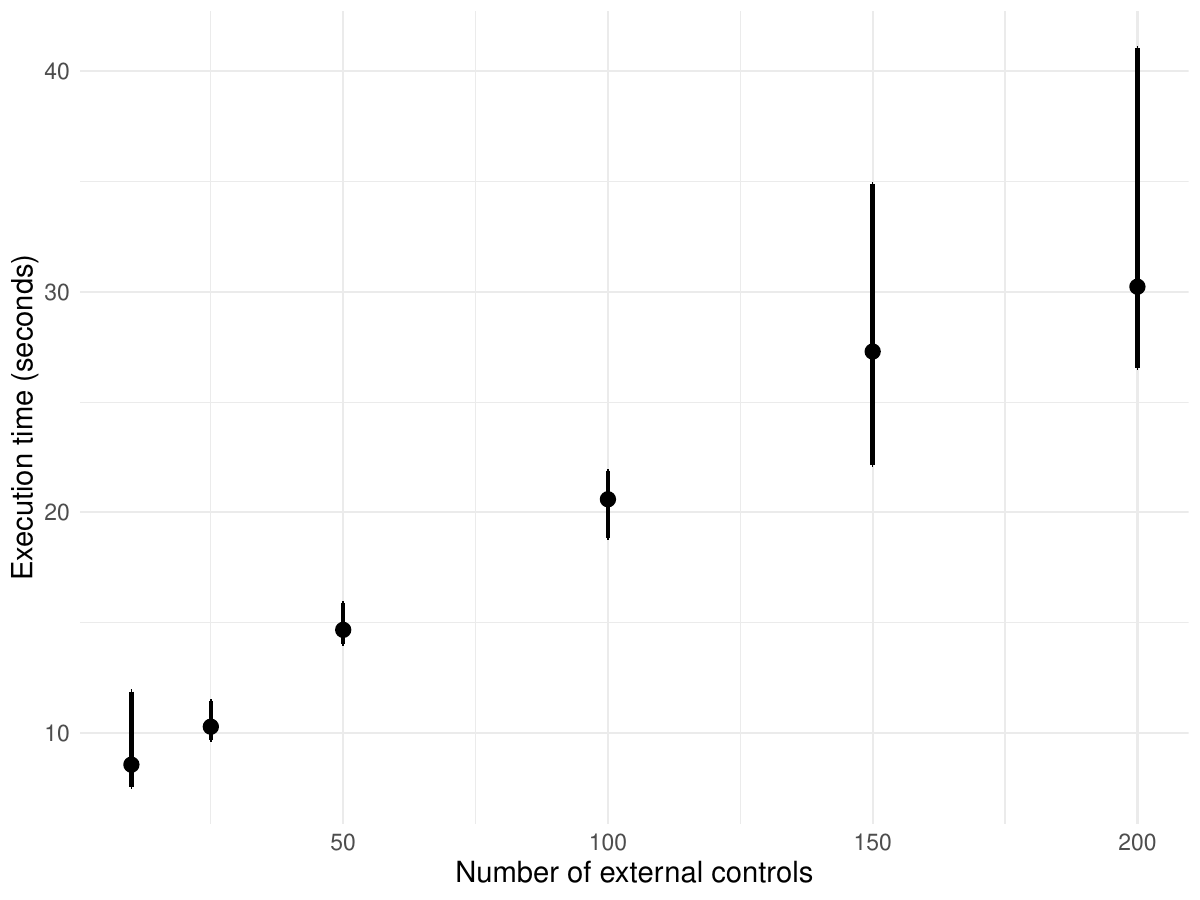}
        \caption{Varying sample size}
        \label{fig:vary_param}
    \end{subfigure}
    \caption{The computational time in seconds to for the proposed combined estimator when varying either the number of external controls or the number of observed covariates, while keeping everything else fixed. Each point corresponds to the median time over 10 repeated runs and the error bars show the min/max time.}
    \label{fig:computational_cost}
\end{figure}

\paragraph{Computational costs with M-estimation} 
We assessed the computational efficiency of our proposed methodology by measuring the execution time of the combined estimator as we increased either the sample size or the number of parameters to be estimated while keeping all other factors constant. In both cases, we observed a linear increase in computational time. The results are shown in Figure~\ref{fig:computational_cost}.

\begin{table}[ht]
\scriptsize
\caption{Simulation study to investigate effect on the optimized randomization-aware estimator when removing the weighting in its objective and another investigation of replacing the true propensity score with an estimated propensity score. We considered a small trial $(n_1=50)$ and a large trial $(n_1=200)$ where the number of external controls was kept fixed $(n_0=200)$. We computed the mean absolute bias, variance and coverage rate from the estimated $95\%$-confidence intervals using 5000 repeated simulations.}
\label{tab:ablation_study}
\begin{tabular*}{\linewidth}{@{\extracolsep{\fill}}crrrrrrrrrrrr}
\toprule
 & \multicolumn{6}{c}{Scenario A} & \multicolumn{6}{c}{Scenario B} \\ 
\cmidrule(lr){2-7} \cmidrule(lr){8-13}  & \multicolumn{3}{c}{Small trial} & \multicolumn{3}{c}{Large trial} & \multicolumn{3}{c}{Small trial} & \multicolumn{3}{c}{Large trial} \\ 
\cmidrule(lr){2-4} \cmidrule(lr){5-7} \cmidrule(lr){8-10} \cmidrule(lr){11-13}
Estimator & Bias & Var & Cov & Bias & Var & Cov & Bias & Var & Cov & Bias & Var & Cov \\ 
\midrule\addlinespace[2.5pt]
Optimized randomization-aware & 0.00 & 0.29 & 0.96 & 0.00 & 0.02 & 0.94 & 0.01 & 0.73 & 0.93 & 0.00 & 0.18 & 0.94 \\ 
Remove weighting in optimization & 0.00 & 0.29 & 0.96 & 0.00 & 0.02 & 0.94 & 0.11 & 0.77 & 0.94 & 0.02 & 0.18 & 0.95 \\ 
Estimate propensity score & 0.00 & 0.31 & 0.97 & 0.00 & 0.02 & 0.94 & 0.02 & 0.80 & 0.91 & 0.00 & 0.18 & 0.94 \\ 
\bottomrule
\end{tabular*}
\end{table}

\subsection{Code}
The code to reproduce the simulation studies is publicly available at \url{https://github.com/RickardKarl/IntegratingExternalControls}.
\section{Data application details}
\subsection{Summary statistics of baseline covariates for different populations}\label{app:TableOne}

\begin{table}[ht]
    \centering
    \caption{Baseline characteristics stratified by source trial and treatment}
    \begin{tabular}{lccc}
        \hline
         Characteristic  &  $S=1, A=1$ & $S=1, A=0$& $S=0, A=0$\\
         & (n=97) & (n=91) & (n=111)\\
         \hline
         PANSS baseline score (mean (SD))& 92.64 (11.76) & 93.09 (11.53) & 94.05 (12.65)\\
         PANSS score at week 6 (mean (SD)) & 77.75 (20.34) & 85.92 (23.12) & 90.67 (24.95)\\
         Age (mean (SD) & 41.94 (10.49) & 42.67 (10.85) & 37.61 (10.92)\\
         Gender (Female=1) (n (\%)) & 31 (32.0) & 21 (23.1) & 35 (31.5)\\
         Race (White=1) (n (\%))& 44 (45.4) & 44 (48.4) & 56 (50.5)\\
         \hline
    \end{tabular}
    \label{app:table:TableOne}
\end{table}

\subsection{Additional analyses}\label{app:application_add}
Furthermore, to empirically examine the methods under different sample sizes of the index trial, we varied the sample size of the controls in the index trial by randomly sampling with replacement a fraction of the available observations. Specifically, of the 91 patients in the control group of the index trial; we sampled 68 ($\sim75\%$), 46 ($\sim50\%$), or 23 ($\sim25\%$) individuals and repeated the analyses described above 100 times. The average of the results over the 100 analyses are shown in Table \ref{tab:ApplicationResults_add} (lines 2-4). With small sample sizes in the control group of the index trial (68, 46, and 23), the point estimates of the pooling estimator deviated from the points estimates of IPW and the AIPW estimator. This showed that the pooling estimator is biased in this data application, mostly because conditions 4 and 5 did not hold. On the other hand, the point estimates of the proposed estimators (randomization-aware and combined estimators) were similar to the point estimates of IPW and AIPW estimators, indicating that the proposed estimators do not generate bias when augmenting the index trial using external controls even if conditions 4 and 5 do not hold. In addition, the combined estimator's standard error is always smaller than the standard error of AIPW estimator; the efficiency improvement is larger with the decreasing sample size of the controls in the index trial.

\begin{table}[ht]
    \centering\scriptsize
    \caption{Estimates and standard errors of the different estimators with various sample sizes of the controls in the index trial (100\%, 75\%, 50\%, 25\% of the original sample size, 91). The results were the averages of 100 runs. The results of the test-then-pool estimator are not shown; they are the same as that of the pooling estimator.}
    \label{tab:ApplicationResults_add}
    \begin{tabular}{lccccccc}\hline  
       &Unadjusted & \shortstack{Pooling}&$\widehat \tau(\widehat g)$  & $\widehat \tau(\widehat h^{*})$ & $\widehat \tau(\widehat \lambda^{*})$ & \shortstack{Selective\\borrowing} &\shortstack{Dynamic\\borrowing}\\\hline
    68 &  -7.910 (13.483) &  -9.534 (2.547) & -7.501 (3.164) &-7.241 (3.241) & -7.482 (3.159) & -7.487 (3.091) & -10.272 (2.707)\\
    46  & -8.134 (15.540) & -9.700 (2.663) & -7.667 (3.635) &-7.304 (3.709) & -7.657 (3.609) & -7.705 (3.592) &-10.698 (2.890)\\
    23 & -8.302 (20.623) & -9.888 (2.840) & -8.053 (4.680) & -7.001 (5.003) & -8.047 (4.544) & -8.301 (4.404) &-11.502 (3.090)\\\hline
 \end{tabular}
\end{table}

\clearpage
\bibliographystyle{plainnat}
\bibliography{references}

\end{appendices}

\end{document}